\documentclass[10pt,letterpaper,english,twocolumn]{revtex4-1}
\usepackage{mathptmx}

\usepackage[T1]{fontenc}
\usepackage[latin9]{inputenc}
\usepackage{xcolor}
\usepackage{pdfcolmk}
\usepackage{amsthm}
\usepackage{amsmath}
\usepackage{amssymb}
\usepackage{graphicx}
\usepackage{esint}
\PassOptionsToPackage{normalem}{ulem}
\usepackage{ulem}

\makeatletter

\pdfpageheight\paperheight
\pdfpagewidth\paperwidth

\providecolor{lyxadded}{rgb}{0,0,1}
\providecolor{lyxdeleted}{rgb}{1,0,0}

\@ifundefined{textcolor}{}
{%
 \definecolor{BLACK}{gray}{0}
 \definecolor{WHITE}{gray}{1}
 \definecolor{RED}{rgb}{1,0,0}
 \definecolor{GREEN}{rgb}{0,1,0}
 \definecolor{BLUE}{rgb}{0,0,1}
 \definecolor{CYAN}{cmyk}{1,0,0,0}
 \definecolor{MAGENTA}{cmyk}{0,1,0,0}
 \definecolor{YELLOW}{cmyk}{0,0,1,0}
 }

\newcommand{\ket}[1]{\left|#1\right>}

\newcommand{\nbep}[0]{^9\mbox{Be}^+}
\usepackage{pslatex}

\makeatother

\usepackage{babel}
\begin{document}
\noindent \textbf{Engineered 2D Ising interactions on a trapped-ion
quantum simulator with hundreds of spins}

\rule[0.5ex]{1\columnwidth}{1pt}

Joseph W. Britton$^{1}$, Brian C. Sawyer$^{1}$, Adam C. Keith$^{2,3}$,
C.-C. Joseph Wang$^{2}$, James K. Freericks$^{2}$, Hermann Uys$^{4}$,
Michael J. Biercuk$^{5}$, John. J. Bollinger$^{1}$

$^{1}$US National Institute of Standards and Technology, Time and
Frequency Division, Boulder, CO 80305

$^{2}$Department of Physics, Georgetown University, Washington, DC
20057

$^{3}$Department of Physics, North Carolina Stawte University, Raleigh,
NC 27695

$^{4}$National Laser Centre, Council for Scientific and Industrial
Research, Pretoria, South Africa

$^{5}$Centre for Engineered Quantum Systems, School of Physics, The
University of Sydney, NSW 2006 Australia

\rule[0.5ex]{1\columnwidth}{1pt}

The presence of long-range quantum spin correlations underlies a variety
of physical phenomena in condensed matter systems, potentially including
high-temperature superconductivity \cite{Anderson1987,Moessner2000}.
However, many properties of exotic strongly correlated spin systems
(e.g., spin liquids) have proved difficult to study, in part because
calculations involving $N$-body entanglement become intractable for
as few as $N\sim30$ particles \cite{Sandvik2010}. Feynman divined
that a quantum simulator \textemdash{} a special-purpose \textquotedbl{}analog\textquotedbl{}
processor built using quantum particles (qubits) \textemdash{} would
be inherently adept at such problems \cite{Feynman1982,Buluta2009}.
In the context of quantum magnetism, a number of experiments have
demonstrated the feasibility of this approach \cite{Trotzky2008,Lin2009,Jo2009fmKetterleScience,Friedenauer2008,Kim2010,Simon2011,Ma2011,Islam2011a,Lanyon2011}.
However, simulations of quantum magnetism allowing controlled, tunable
interactions between spins localized on 2D and 3D lattices of more
than a few 10's of qubits have yet to be demonstrated, owing in part
to the technical challenge of realizing large-scale qubit arrays.
Here we demonstrate a variable-range Ising-type spin-spin interaction
$J_{i,j}$ on a \textit{naturally occurring} 2D triangular crystal
lattice of hundreds of spin-1/2 particles ($\nbep$ ions stored in
a Penning trap), a computationally relevant scale more than an order
of magnitude larger than existing experiments. We show that a spin-dependent
optical dipole force can produce an antiferromagnetic interaction
$J_{i,j}\propto d_{i,j}^{-a}$, where $a$ is tunable over $0<a<3$;
$d_{i,j}$ is the distance between spin pairs. These power-laws correspond
physically to infinite-range ($a=0$), Coulomb-like ($a=1$), monopole-dipole
($a=2$) and dipole-dipole ($a=3$) couplings. Experimentally, we
demonstrate excellent agreement with theory for $0.05\lesssim a\lesssim1.4$.
This demonstration coupled with the high spin-count, excellent quantum
control and low technical complexity of the Penning trap brings within
reach simulation of interesting and otherwise computationally intractable
problems in quantum magnetism. 

\rule[0.5ex]{1\columnwidth}{1pt}

A challenge in condensed matter physics is the fact that many quantum-magnetic
interactions cannot currently be modeled in a meaningful way. A canonical
example is the spin-liquid postulated by Anderson \cite{Anderson1987}.
He suggested this exotic state would arise in a collection of spin-1/2
particles residing on a triangular lattice and coupled by a nearest-neighbor
antiferromagnetic Heisenberg interaction. The spin-liquid's ground
state is massively degenerate, owing to spin frustration, and is expected
to exhibit unusual behaviors including phase transitions at zero temperature
driven by quantum fluctuations \cite{Sachdev2001}. However, despite
recent advances \cite{Kohno2007,Varney2011} a detailed understanding
of large-scale frustration in solids remains elusive \cite{Moessner2000,Levi2007,Helton2007,Balents2010}. 

\begin{figure}[t]
\includegraphics[width=3.5in]{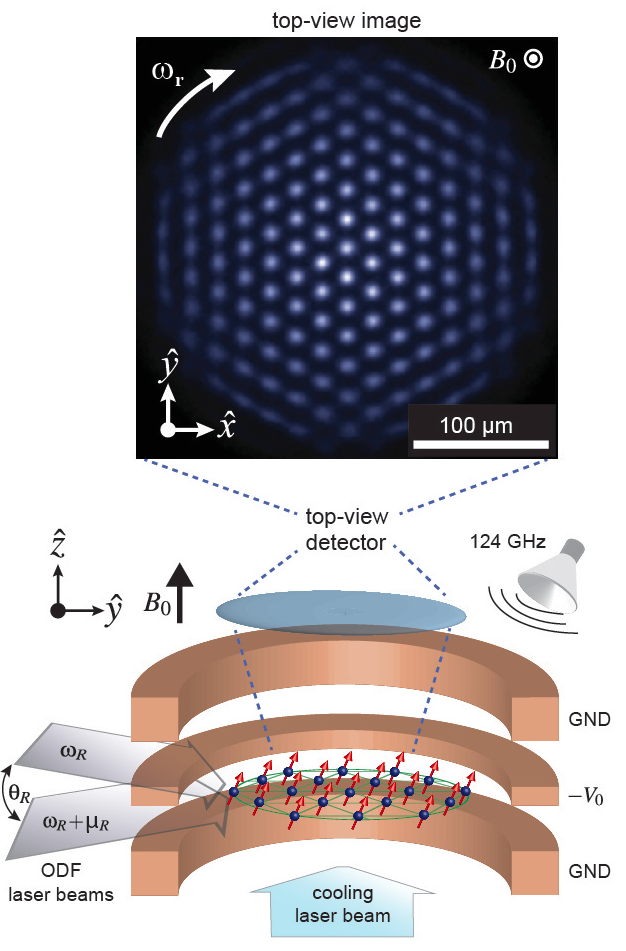}

\caption{The Penning trap confines hundreds of spin-$1/2$ particles (qubits)
on a two-dimensional (2D) triangular lattice. Each qubit is the valence
electron spin of a $\nbep$ ion. (lower) A Penning trap confines ions
by use of a combination of static electric and magnetic fields. The
trap parameters are configured so that laser-cooled ions form a triangular
2D crystal. A general spin-spin interaction $\hat{H}_{I}$ is generated
by a spin-dependent excitation of the transverse (along~$\hat{z}$)
motional modes of the ion crystal. This coupling is implemented with
an optical dipole force (ODF) due to a pair of off-resonance laser
beams (left side) with angular separation $\theta_{R}$ and difference
frequency $\mu_{R}$. Microwaves at $124\,\mbox{GHz}$ are directed
to the ions with a horn and permit global spin rotations $\hat{H}_{B}$.
(upper) A representative top-view resonance-fluorescence image showing
the center region of an ion crystal captured in the ions' rest-frame;
in the lab-frame the ions rotate a at $\omega_{r}$$=2\pi\times43.8$~kHz
\cite{Mitchell1998}. Fluorescence is an indication of the qubit spin-state
($\ket{\uparrow}$ bright, $\ket{\downarrow}$ dark); here, the ions
are in the state $\ket{\uparrow}$. The lattice constant is $d_{0}\sim20\,\mu\mbox{m}$. }

\label{Fig:blingView} \label{figTopView} \label{figRamanGeometry}
\end{figure}

Atomic physicists have recently entered the fray, providing a bottom-up
approach by engineering the relevant spin interactions in quantum
simulators \cite{Lewenstein2007,Bloch2008,Buluta2009}. The necessary
experimental capabilities were first demonstrated in the context of
atomic clocks: laser cooling, deterministic spin localization, precise
spin-state quantum control, high-fidelity readout, and engineered
spin-spin coupling (e.g.,~\cite{Rosenband2008}). In the domain of
quantum magnetism, this tool set permits control of parameters commonly
viewed as immutable in natural solid systems, e.g., lattice spacing
and geometry, and spin-spin interaction strength and range. 

Initial simulations of quantum Ising and Heisenberg interactions with
localized spins were done with neutral atoms in optical lattices \cite{Trotzky2008,Simon2011},
atomic ions in Paul traps \cite{Friedenauer2008,Kim2010,Islam2011a,Lanyon2011},
and photons \cite{Ma2011}. This work involved a domain readily calculable
on a classical computer: interactions between $N\sim10$ qubits localized
in 1D chains. The move to quantum magnetic interactions on 2D lattices
and larger, computationally relevant particle numbers is the crucial
next step but can require significant technological development \cite{Schmied2009}. 

In our Penning trap apparatus, laser-cooled $\nbep$ ions \textit{naturally}
form a stable 2D Coulomb crystal on a triangular lattice with $\sim300$
spins (Fig.~\ref{Fig:blingView}). Each ion is spin-1/2 system (qubit)
over which we exert high fidelity quantum control \cite{Biercuk2009ff}.
In this paper we demonstrate the use of a spin-dependent optical dipole
force (ODF) to engineer a continuously-tunable Ising-type spin-spin
coupling $J_{i,j}\propto d_{i,j}^{-a}$. This capability, in tandem
with a modified measurement routine (e.g., by more sophisticated processing
of images like Fig.~\ref{Fig:blingView}), is a key advance toward
useful simulations of quantum magnetism. 

\begin{figure}
\includegraphics[width=3.5in]{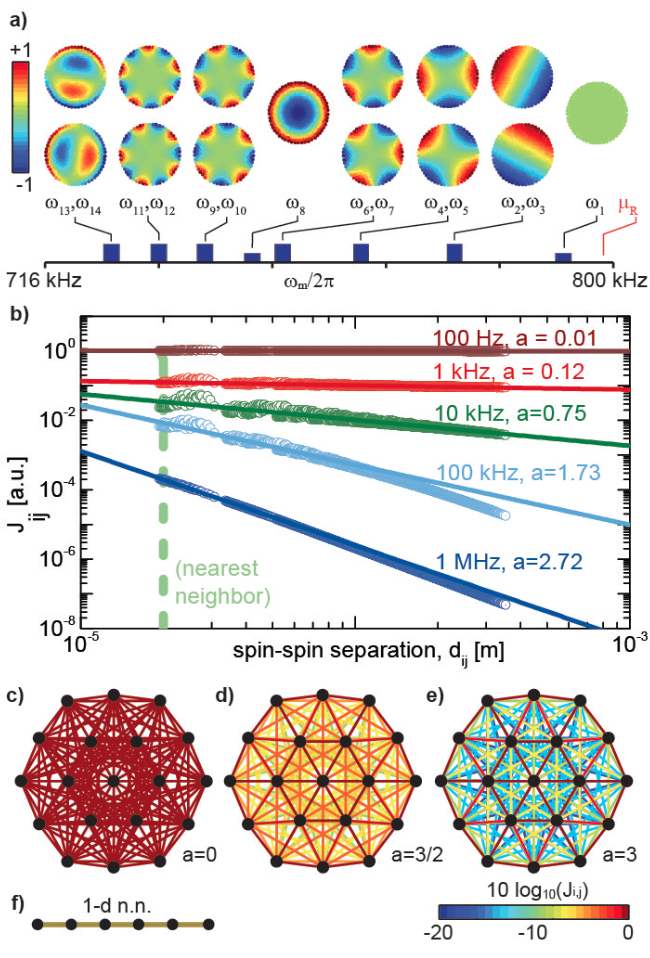}\caption{Spin-spin interactions are mediated by the ion crystal's transverse
motional degrees of freedom. (a) For a 2D crystal with $N=217$ ions
and $\omega_{r}=2\pi\times45.6$~kHz, we calculate the eigenfunctions
$\vec{b}_{m}$ and eigenfrequencies $\omega_{m}$ for the $N$ transverse
motional modes (see Supplementary Information). Plotted here are $\omega_{m}$
and $\vec{b}_{m}$ for the $14$ highest frequency modes. Relative
mode amplitude is indicated by color. The center-of-mass motion (COM)
is the highest in frequency ($\omega_{1}\sim2\pi\times795$~kHz)
and has no spatial variation in $\vec{b}_{m}$. The lowest-frequency
mode ($\omega_{217}\sim2\pi\times200$~kHz) exhibits spatial variation
in $\vec{b}_{m}$ at the lattice spacing length scale $d_{0}\sim20\,\mu\mbox{m}$.
(b) By use of Eq.~\ref{eq:Jij}, $J_{i,j}$ is calculated explicitly
for $N=217$ spins and plotted as a function of spin-spin separation
$d_{i,j}$. For $\mu_{R}-\omega_{1}<2\pi\times1$~kHz, $\hat{H}_{ODF}$
principally excites COM motion in which all ions equally participate:
the spin-spin interaction is spatially uniform. As the detuning is
increased, modes of higher spatial frequencies participate in the
interaction and $J_{i,j}$ develops a finite interaction length. We
find the scaling of $J_{i,j}$ with spin separation $d_{i,j}$ follows
the power law $d_{i,j}^{-a}$. For $\mu_{R}-\omega_{1}\gg2\pi\times500\,\mbox{kHz}$,
all transverse modes participate and the spin-spin coupling power
law approaches $a=3$. The solid lines are power-law fits to the theory
points. For comparison with other experiments, the nearest neighbor
coupling ($d_{0}=20\,\mu\mbox{m}$) is denoted by the dashed line.
(c)-(e) The power-law nature of $J_{i,j}$ is qualitatively illustrated
for $N=19$ (larger $N$ is illegible). Nodes (spins) are joined by
lines colored in proportion to their coupling strength for various
$a$. (f) For context, the graph for a 1D nearest-neighbor Ising interaction,
a well-known model in quantum field theory. }

\label{Fig:Jijtheory} \label{Fig:theoryEigenmodes} 
\end{figure}

A Penning trap confines ions in a static quadrupolar electric potential
(see Methods) and a strong, homogeneous magnetic field $B_{0}\hat{z}$
(here, $B_{0}=4.46$~Tesla). Axial trapping (along $\hat{z}$) is
due to the electric field. Ion rotation at frequency $\omega_{r}$
(about $\hat{z}$) produces a radial restoring potential due to the
velocity-dependent Lorentz force ($q\vec{v}\times\mbox{\ensuremath{\vec{B}}}$).
Tuning the ratio of the axial to radial confinement allows controlled
formation of a planar geometry and, after Doppler laser cooling, the
formation of a 2D Coulomb crystal on a triangular lattice (see \cite{Mitchell1998}
and Methods). We routinely generate crystals with $100\lesssim N\lesssim300$
ions, where the valence-electron spin state of each ion serves as
a qubit \cite{Biercuk2009ff}. Following techniques developed in linear
(1D) Paul traps \cite{Leibfried2003}, spins confined in the same
trapping potential are coupled through their shared motional degrees
of freedom.

Using well controlled external fields, we engineer spin interactions
of the form 

\begin{equation}
\begin{array}{rcl}
\hat{H}_{B} & = & \sum_{i}\vec{B_{\mu}}\cdot\hat{\vec{\sigma}}_{i}\\
\hat{H}_{I} & = & \frac{1}{N}\sum_{i<j}J_{i,j}\hat{\sigma}_{i}^{z}\hat{\sigma}_{j}^{z},
\end{array}\label{eq:HB_HI}
\end{equation}
where $\hat{\sigma}_{i}^{z}$ is the $z$-Pauli matrix for ion $i$.
We label the qubit spin states $\ket{\uparrow}\equiv\ket{m_{s}=+1/2}$
and $\ket{\downarrow}\equiv\ket{m_{s}=-1/2}$, where $m_{s}$ is the
spin's projection along the $B_{0}\hat{z}$ quantizing field, so that
$\hat{\sigma}_{i}^{z}\ket{\uparrow_{i}}=+\ket{\uparrow_{i}}$ and
$\hat{\sigma}_{i}^{z}\ket{\downarrow_{i}}=-\ket{\downarrow_{i}}$.
$\hat{H}_{B}$ is an interaction (e.g., $B_{\mu}\hat{\sigma}_{x}$)
generated by externally applied microwaves at 124~GHz that couples
equally to all spins and permits global rotations (Fig.~\ref{Fig:blingView}).
The interaction $\hat{H}_{I}$ describes a general coupling $J_{i,j}$
between spins $i$ and $j$, distance $d_{i,j}$ apart \cite{Porras2006,Kim2009};
$J_{i,j}>0$ ($J_{i,j}<0$ ) is an antiferromagnetic (ferromagnetic)
coupling. 

We implement $\hat{H}_{I}$ using a spatially uniform spin-dependent
optical dipole force (ODF) generated by a pair of off-resonant laser
beams with difference frequency $\mu_{R}$ (see Fig.~\ref{figRamanGeometry}
and Supplementary Information). The ODF couples each ion's spin to
one or more of the $N$~transverse (along $\hat{z}$) motional modes
of the Coulomb crystal by forcing coherent displacements of the ions
that in turn modify their Coulomb potential energy through the interaction
\begin{equation}
\hat{H}_{ODF}=-\sum_{i}^{N}F_{z}(t)\hat{z}_{i}\hat{\sigma}_{i}^{z}.\label{eq:HODF}
\end{equation}
Here $F_{z}(t)=F_{0}\cos\mu_{R}t$ is the ODF, $\hat{z}_{i}=\sum_{m=1}^{N}b_{i,m}\sqrt{\frac{\hbar}{2M\omega_{m}}}\left(\hat{a}_{m}e^{-i\omega_{m}t}+\hat{a}_{m}^{\dagger}e^{i\omega_{m}t}\right)$
is the axial position operator for ion $i$, and $b_{i,m}$ are elements
of the $N$ transverse eigenfunctions $\vec{b}_{m}$ at frequencies
$\omega_{m}$ normalized as $\sum_{m=1}^{N}|b_{i,m}|^{2}=\sum_{i=1}^{N}|b_{i,m}|^{2}=1$
\cite{Porras2006,Kim2009}. The modes include the center-of-mass mode
($\omega_{1}$) as well as an array of higher spatial-frequency modes
that may be derived from atomistic calculations (see Fig.~\ref{Fig:theoryEigenmodes})
and confirmed by experimental measurement \cite{Sawyer2012modeAndTempSpec}. 

For small coherent displacements, where residual spin-motion entanglement
can be neglected (see \cite{Kim2009} and Methods), $\hat{H}_{ODF}$
is equivalent to $\hat{H}_{I}$ in Eq.~\ref{eq:HB_HI}: spins $i$~and~$j$
are coupled in proportion to their spin states $\hat{\sigma}_{i}^{z}$
and $\hat{\sigma}_{j}^{z}$ and mutual participation in each motional
mode $m$. The coupling coefficient is given by \cite{Kim2009} 
\begin{equation}
J_{i,j}=\frac{F_{0}^{2}N}{2\hbar M}\sum_{m=1}^{N}\frac{b_{i,m}b_{j,m}}{\mu_{R}^{2}-\omega_{m}^{2}}.\label{eq:Jij}
\end{equation}
These pairwise interaction coefficients $J_{i,j}$ can be calculated
explicitly by use of Eq.~\ref{eq:Jij} and classical calculations
of ion motional modes. We find that the range of interaction can be
modified by detuning away from the COM mode as discussed in Fig.~\ref{Fig:Jijtheory}.
In the limit $\mu_{R}-\omega_{1}\gg2\pi\times500$~kHz, all modes
participate equally in the interaction and $J_{i,j}\propto d_{i,j}^{-3}$,
as discussed in \cite{Porras2006}. At intermediate detuning, we find
a power-law scaling of the interaction range $J_{i,j}\propto d_{i,j}^{-a}$,
where $a$ can be tuned $0\leq a\leq3$. That is, by adjusting the
single experimental parameter $\mu_{R}$ we can mimic a continuum
of physical couplings including important special cases: $a=0$ is
infinite-range, $a=1$ is monopole-monopole (Coulomb-like), $a=2$
is monopole-dipole and $a=3$ is dipole-dipole. Note that $a=0$ results
in the so-called $\hat{J}_{z}^{2}$ interaction that gives rise to
spin-squeezing and is used in quantum logic gates (see Supplementary
Information) \cite{Leibfried2003}. In addition, tuning $\mu_{R}$
also permits access to both antiferromagnetic (AFM, $\mu_{R}>\omega_{1}$)
and ferromagnetic (FM, $\omega_{2}\ll\mu_{R}<\omega_{1}$) couplings
\cite{Islam2011a}. 

\begin{figure}
\includegraphics[width=3.15in]{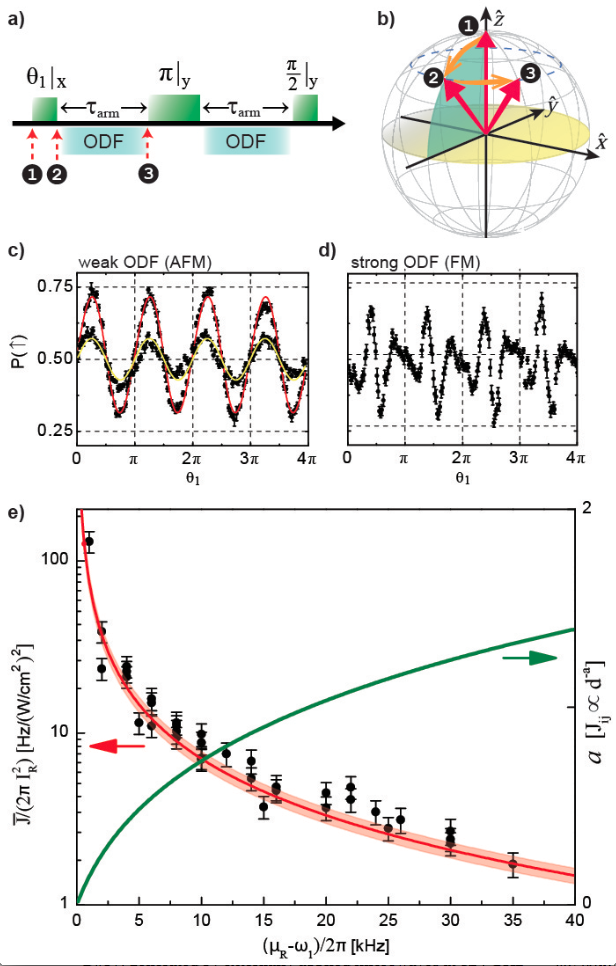}

\caption{Benchmarking the 2D Ising interaction. (a) Spin-precession benchmarking
sequence for $\hat{H}_{I}$. The spins are prepared at the outset
in $\ket{\uparrow}$ (a ferromagnetic state). The spin-spin interaction
$\hat{H}_{I}$ is present when the ODF laser beams are on. We choose
$\mu_{R}-\omega_{1}=n\times2\pi/\tau_{arm}$ so that for small detunings
from the COM mode ($\omega_{1}$), the spins are decoupled from the
motion at the end of the drive. (b) Evolution of a single spin prior
to the application of the spin-echo $\pi$-pulse. c,d) Plots exhibiting
spin precession $\propto\left<\hat{\sigma}_{z}\right>$ due to $\hat{H}_{I}$
as a function of initial tipping angle $\theta_{1}$. The error bars
are statistical (s.d., $n\sim200$). The plots show typical experimental
data (black) and single-parameter fits to Eq.\ref{eq:chi_vs_theta1_vs_gamma}.
For an antiferromagnetic coupling (AFM) and $\tau_{arm}=250\,\mu\mbox{s}$,
$\mu_{R}-\omega_{1}=2\pi\times4.0$~kHz, $I_{R}=1.4\,\mbox{W\ensuremath{\cdot}cm}^{-2}$
we obtain $\bar{J}/I_{R}^{2}=2\pi\times25$~$\mbox{Hz}\cdot\mbox{W}^{-2}\cdot\mbox{cm}^{4}$
(yellow fit). Longer drive periods and higher laser intensity $I_{R}$
yield a larger precession. For $\tau_{arm}=350\,\mu\mbox{s}$, $\mu_{R}-\omega_{1}=2\pi\times2.9$~kHz,
$I_{R}=1.9\,\mbox{W\ensuremath{\cdot}cm}^{-2}$ we obtain $\bar{J}/I_{R}^{2}=2\pi\times55$~$\mbox{Hz}\cdot\mbox{W}^{-2}\cdot\mbox{cm}^{4}$
(red fit). The data in this plot is typical of the experiments conducted
for benchmarking. (d) For a much stronger interaction, Eq.~\ref{eq:chi_vs_theta1_vs_gamma}
cannot be used to obtain $\bar{J}$ since the MF assumption is no
longer valid (see Supplementary Information). Also, here we used a
small negative detuning ($\omega_{2}\ll\mu_{R}<\omega_{1}$) which
gives a long-range ferromagnetic (FM) interaction. For these experiments
we set $\omega_{r}=2\pi\times45.6$~kHz. (e) Benchmarking results
for an ion-crystal with $N=206\pm10$ ions. Each point is generated
by measuring $\bar{J}$ as in c) and measuring the laser beam intensity
at the ions $I_{R}$. The vertical error bars are dominated by uncertainty
in $I_{R}$ (see Supplementary Information). The solid line (red)
is the prediction of MF theory that accounts for couplings to all
$N$ transverse modes; there are no free parameters. The line's breadth
reflects experimental uncertainty in the angle $\theta_{R}=4.8^{\circ}\pm0.25$.
The MF prediction for the average power-law scaling coefficient $a$
is drawn in green (right axis, linear scale). }

\label{Fig:expSequence} \label{Fig:JzFlopping} \label{Fig:benchmarking}
\end{figure}

Experimentally, we demonstrate a tunable-range Ising interaction by
observing a global spin precession under application of $\hat{H}_{I}$.
We compare experimental data with the mean field (MF) prediction that
the influence of $\hat{H}_{I}$ on spin $j$ can be modeled as an
excess magnetic field in the $\hat{z}$-direction $\bar{B}_{j}$ due
to the remaining $N-1$ spins, $\bar{B}_{j}=\frac{2}{N}\sum_{i,i\neq j}^{N}J_{i,j}\left<\hat{\sigma}_{i}^{z}\right>$
(see Supplementary Information). For a general qubit superposition
state, $\bar{B}_{j}$ gives rise to spin precession about $\hat{z}$
in excess of that expected due to simple Larmor precession. The experiment
sequence shown in Fig.~\ref{Fig:expSequence}a measures this excess
precession, averaged over all spins in the crystal. At the outset,
each spin is prepared in $\left|\uparrow\right\rangle $, followed
by a rotation about the $\hat{x}$-axis by an angle $\theta_{1}$.
$\hat{H}_{I}$ is applied during the arms of a spin-echo; precession
$\propto\left<\hat{\sigma}_{i}^{z}\right>$ coherently adds over both
spin-echo intervals. The final $\pi/2$-pulse maps precession out
of the initial $\widehat{yz}$-plane into excursions along $\hat{z}$
(above or below the equatorial plane of the Bloch sphere) that are
resolved by projective spin measurement along $\hat{z}$. 

As a function of ``tipping-angle'' $\theta_{1}$ we detect global
state-dependent fluorescence ($\ket{\uparrow}$ bright, $\ket{\downarrow}$
dark) using a photomultiplier tube. This measurement permits a systematic
study of MF-induced spin precession averaged over all particles, $\frac{1}{N}\sum_{j}^{N}\bar{B}_{j}=2\left(\frac{1}{N^{2}}\sum_{j}^{N}\sum_{i,i\neq j}^{N}J_{i,j}\right)\cos\theta_{1}\equiv2\bar{J}\cos\theta_{1}$.
The probability of detecting $\ket{\uparrow}$ at the end of the sequence
is 

\begin{equation}
P(\ket{\uparrow})=\frac{1}{2}\left(1+\exp(-\Gamma\cdot2\tau_{arm})\sin(\theta_{1})\sin(2\bar{J}\cos(\theta_{1})\cdot2\tau_{arm})\right),\label{eq:chi_vs_theta1_vs_gamma}
\end{equation}
and a single-parameter fit to experimental data yields $\bar{J}$.
Decoherence due to spontaneous emission is accounted for by $\Gamma$
and is fixed by independent measurement of the ODF laser beam intensities
$I_{R}$ ($\Gamma\propto I_{R}$, see Supplementary Information). 

In Fig.~\ref{Fig:JzFlopping}c,d we show representative measurements
of excess precession due to $\hat{H}_{I}$ for different values of
spin coupling strength (determined by $J_{i,j}\propto I_{R}^{2}$)
and interaction duration $2\tau_{arm}$. The excess spin precession
varies periodically with $\theta_{1}$ (period $\pi$) and larger
interaction strength yields greater precession, manifested in our
experiment as a larger amplitude modulation of $P(\ket{\uparrow})$.
Data agree with Eq.~\ref{eq:chi_vs_theta1_vs_gamma} and allow direct
extraction of $\bar{J}$ for given experimental conditions. In Fig.~\ref{Fig:benchmarking}e
we plot $\bar{J}$, normalized by $I_{R}^{2}$ ($I_{R}$ independently
measured), as function of detuning $\mu_{R}-\omega_{1}$ ($N=206\pm10$
ions). Using no free parameters, we find excellent agreement with
$\bar{J}$ obtained by averaging over all $J_{i,j}$, where the $J_{i,j}$
were calculated by including couplings to all $N$ transverse modes
(Eq.~\ref{eq:Jij}). 

The MF interpretation of our benchmarking measurement tolerates only
weak spin-spin correlations. Therefore, in the benchmarking regime
we apply a weak interaction ($\bar{J}\cdot2\tau_{arm}\ll\sqrt{N}/4$,
see Supplementary Information). In a quantum simulation, the same
interaction is applied at greater power producing quantum spin-spin
correlations. In the present configuration of our apparatus, spontaneous
emission due to the ODF laser beams ($\Gamma$ in Eq.~\ref{eq:chi_vs_theta1_vs_gamma})
is the dominant source of decoherence. With modest laser powers of
$4\,\mbox{mW}$/beam and a detuning $\mu_{R}-\omega_{1}=2\pi\times2\,\mbox{kHz}$,
we obtain $\bar{J}\sim2\pi\times0.5\,\mbox{kHz}$ and $\Gamma/\bar{J}\sim0.06$.
The expected spin-squeezing ($\hat{J}_{z}^{2}$) due to this interaction
is $5\,\mbox{dB}$, limited by spontaneous emission. The ratio $\Gamma/\bar{J}$
can be reduced ($\bar{J}$ increased) by a factor of $50$ by increasing
$\theta_{R}$ to $35^{\circ}$, a likely prerequisite for access to
the shortest-range, dipole-dipole coupling regime ($a\rightarrow3$).
At present, geometric constraints limit $\theta_{R}\lesssim5^{\circ}$;
we plan upgrades to our apparatus to permit $\theta_{R}=35^{\circ}$.
We also note that relaxation of the constraint $F_{\uparrow}\sim-F_{\downarrow}$
can also reduce $\Gamma/\bar{J}$. 

In summary, this work establishes the suitability of a Penning trap
apparatus to pursue quantum simulation in a regime inaccessible to
classical computation. Our approach is based on naturally occurring
2D Coulomb (Wigner) crystals with hundreds of ion qubits, a novel
experimental system that does not require demanding trap-engineering
efforts. Experimentally, we used an optical-dipole force to engineer
a tunable-range spin-spin interaction and benchmarked its interaction
strength. Excellent agreement is obtained with the predictions of
mean field theory and atomistic calculations that predict a power-law
antiferromegnetic spin coupling $J_{i,j}\propto d_{i,j}^{-a}$ for
$0.05\lesssim a\lesssim1.4$. 

With this work as a foundation, we anticipate a variety of future
investigations. For example, simultaneous application of non-commuting
interactions $\hat{H}_{B}$ and $\hat{H}_{I}$ is expected to give
rise to quantum phase transitions; $\hat{H}_{I}$ may be antiferromagnetic
($\mu_{R}>\omega_{1}$) or ferromagnetic ($\omega_{2}\ll\mu_{R}<\omega_{1}$)).
Geometric modifications to our apparatus will permit access to larger
$\theta_{R}$ and antiferromagnetic dipole-dipole-type couplings ($a\rightarrow3$).
Improved image processing software will permit direct measurement
of spin-spin correlation functions using our existing single-spin-resolving
imaging system (Fig.~\ref{figTopView}).

\section*{Methods}

In a frame rotating at frequency $\omega_{r}$, the trap potential
is  
\begin{equation}
q\phi(r,z)=\frac{1}{2}M\omega_{1}^{2}(z^{2}+\beta r^{2}),\label{eq:trapPotential}
\end{equation}
where $q$ is the ion charge, $M$ is the single-ion mass and $\beta=\omega_{r}\omega_{1}^{-2}(\Omega_{c}-\omega_{r})-1/2$.
The $\nbep$ cyclotron frequency is $\Omega_{c}=B_{0}q/M=2\pi\times7.6$~MHz
and the frequency of the ions' harmonic center-of-mass (COM) motion
along $\hat{z}$ is $\omega_{1}=2\pi\times795$~kHz. Ion rotation
is precisely controlled with an external rotating quadrupole potential
\cite{Mitchell1998}. For $100\lesssim N\lesssim300$ we set $\omega_{r}\sim2\pi\times45$~kHz
so that the radial confinement is sufficiently weak that a cloud of
ions relaxes into a single 2D plane ($\beta\ll1$). Upon Doppler laser
cooling the ions' motional degrees of freedom ($T_{COM}\sim1$~mK)
\cite{Sawyer2012modeAndTempSpec}, the ions naturally form a 2D Coulomb
crystal on a triangular lattice, the geometry that minimizes the energy
of their mutual Coulomb potential energy. The crystal has $N$ transverse
eigenmodes $\omega_{m}$ with eigenfunctions $\vec{b}_{m}$; the COM
mode $\omega_{1}$ is the highest frequency mode (see Fig.~\ref{Fig:theoryEigenmodes}).

The spin-dependent ODF is generated by a pair of off-resonance laser
beams with angular separation $\theta_{R}\sim4.8^{\circ}$ and difference
frequency $\mu_{R}$ (see Fig.~\ref{Fig:blingView}). The result
is a traveling 1D optical lattice of wavelength $\lambda_{R}=2\pi/|\overrightarrow{\Delta k}|\approx3.7\,\mu\mbox{m}$
whose wavefronts propagate along $\mbox{\ensuremath{\hat{z}}}$, traversing
the ion crystal at frequency $\mu_{R}/2\pi$. Alignment of $\overrightarrow{\Delta k}$$ $
is crucial for proper spin-spin coupling (see Supplementary Information).
The lattice's polarization gradient induces a differential AC Stark
shift on the qubit states (a spin-dependent force). We choose operating
conditions that give $F_{\uparrow}\approx-F_{\downarrow}$, where
$F_{\uparrow}=F_{0}\cos(\mu_{R}t)\hat{z}$. For reference, if the
single-beam intensity at the ions is $I_{R}=1\,\mbox{W}\cdot\mbox{cm}^{-2}$,
we obtain $F_{0}\sim1.4\times10^{-23}$~N. 

Small coherent displacements that produce negligible spin-motion entanglement
(as required by Eq.~\ref{eq:Jij}) are obtained for detunings satisfying
\begin{equation}
\hbar\left|\mu_{R}-\omega_{m}\right|>F_{0}\sqrt{\hbar(2\bar{n}_{m}+1)/(2M\omega_{m})}.\label{eq:spin-motion-detuning}
\end{equation}
This is a more stringent criterion than that used by others \cite{Kim2009,Islam2011a}
as it includes an additional $\sqrt{N}$ to account for a typical
distribution of composite spin states. Moreover, we also include a
correction factor for finite temperature $\bar{n}_{m}\sim k_{B}T/\hbar\omega_{m}$.

\section*{Acknowledgments}

This work was supported by the DARPA OLE programme and NIST. A.C.K.
was supported by the NSF under grant number DMR-1004268. B.C.S. is
supported by an NRC fellowship funded by NIST. J.K.F. was supported
by the McDevitt endowment bequest at Georgetown University. M.J.B.
and J.J.B. acknowledge partial support from the Australian Research
Council Center of Excellence for Engineered Quantum Systems CE110001013.
We thank F. Da Silva, R. Jordens, D. Leibfried, A. O\textquoteright{}Brien,
R. Scalettar and A. M. Rey for discussions.

\section*{Distribution Information}

This manuscript is a contribution of the US National Institute of
Standards and Technology and is not subject to US copyright. Correspondence
and requests for materials should be addressed to J.W.B. (joe.britton@gmail.com).
This version of the manuscript was generated by J.W.B. on April 25,
2012. \clearpage{}

\section*{Supplementary Information}

A review of ion confinement in Penning traps and discussion of a variety
of equilibrium states, including 2D Coulomb crystals (ionic Wigner
crystals), can be found in \cite{Brewer1988,Mitchell1998,Dubin1999}.
A number of authors have theoretically analyzed and discussed the
prospects of using 2D Coulomb crystals for quantum information and
computation \cite{Porras2004,Porras2006,Buluta2008,Taylor2008,Zou2010,Baltrusch2011}.
The engineered Ising interaction, which we report here, builds on
our previous experimental work using 2D Coulomb crystals for high-fidelity
quantum control \cite{Biercuk2009a,Biercuk2009ff,Biercuk2009c,Uys2009,Uys2010}.
Below we discuss some details of this new capability.

\section{Spin initialization, control, and measurement}

\label{sec:qubitControlAndReadout}

\begin{figure}[b]
\begin{centering}
\includegraphics[width=0.95\columnwidth]{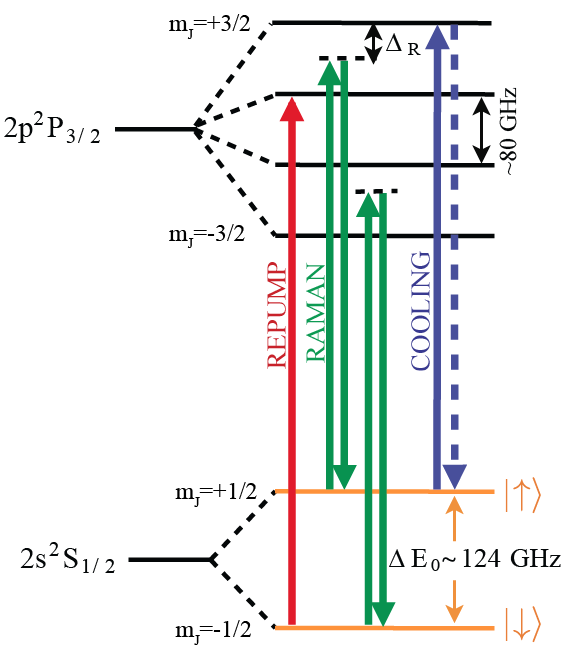}
\par\end{centering}

\caption{Relevant energy levels of $^{9}$Be$^{+}$ at $B_{0}=4.46$~T (not
drawn to scale). We show only $m_{I}=+\frac{3}{2}$ levels that are
prepared experimentally through optical pumping. The $^{2}S_{1/2}-{}^{2}P_{3/2}$
transition wavelength is $\sim313$~nm. A resonant laser beam provides
Doppler laser cooling and state discrimination; a second laser beam
repumps $\left|\downarrow\right\rangle $ to $\left|\uparrow\right\rangle $.
The optical dipole force (ODF) interaction is due to a pair of beams
(derived from the same laser) with relative detuning $\mu_{R}$. The
qubit splitting $\Delta E_{0}/\hbar\sim2\pi\times124$~GHz. A low-phase-noise
microwave source at $124$~GHz provides full global control over
spins. }
\label{figLevelDiagram} 
\end{figure}

Reference \cite{Biercuk2009ff} gives a detailed discussion of our
spin initialization, control, and measurement capabilities with planar
ion arrays in Penning traps. Here we briefly summarize some of that
discussion, emphasizing aspects relevant for the measurements reported
here. Figure~\ref{figLevelDiagram} shows the relevant $\nbep$ energy
levels. We use the valence electron spin states parallel $\left|\uparrow\right\rangle =\left|m_{J}=+\frac{1}{2}\right\rangle $
and antiparallel $\left|\downarrow\right\rangle =\left|m_{J}=-\frac{1}{2}\right\rangle $
to the applied magnetic field of the Penning trap as the two-level
system or qubit. In the $4.46$~T magnetic field of the trap, these
levels are split by approximately $\Omega_{0}=2\pi\times124$~GHz.
The $\nbep$ nucleus has spin $I=3/2$. However, we optically pump
the nuclear spin to the $m_{I}=+3/2$ level \cite{Itano1981}, where
it remains throughout the duration of an experiment. The ions are
Doppler laser-cooled to a temperature $\sim1$~mK \cite{Sawyer2012modeAndTempSpec}
by a $1$~MHz linewidth, $313$~nm laser tuned approximately 10
MHz below the $\left|\uparrow\right\rangle \rightarrow\left|^{2}P_{3/2}\: m_{J}=+3/2\right\rangle $
cycling transition. Spins in the $\left|\downarrow\right\rangle $
state are efficiently optically pumped to the $\left|\uparrow\right\rangle $
state by a laser tuned to the $\left|\downarrow\right\rangle \rightarrow\left|^{2}P_{3/2}\: m_{J}=+1/2\right\rangle $
transition. The repump beam and the main Doppler laser cooling beam
are directed along the magnetic field ($\hat{z}$-axis). Powers are
a few milli-Watt with laser beam waists of $\sim1$~mm. In addition,
a weak Doppler laser cooling beam ($\sim40\:\mu$m waist) directed
perpendicularly to the $\hat{z}$-axis directly Doppler cools the
perpendicular degrees of freedom. A typical experimental cycle starts
with $10$~ms to $20$~ms of combined Doppler laser cooling and
repumping. The repump laser remains on for another $3$~ms after
the Doppler cooling laser is turned off. The fidelity of the $\left|\uparrow\right\rangle $
state preparation is estimated to be very high ($\gg99.9\%$)~\cite{Biercuk2009ff}. 

Low-phase-noise microwave radiation at 124 GHz is used to rotate the
spins through the magnetic dipole interaction $\hat{H}_{B}=g\mu_{B}\sum_{i}\vec{B}_{\mu}(t)\cdot\left(\vec{\hat{\sigma}}_{i}/2\right)\:$,
where $\vec{B}_{\mu}(t)$ is the applied microwave field (predominantly
perpendicular to $\hat{z}$), $g\simeq2$ is the electron g-factor,
and $\mu_{B}$ is the Bohr magnetron. The fidelity of a $\pi$-pulse
was measured to be greater than $99.9\%$ in a random benchmarking
experiment \cite{Biercuk2009ff}. The microwave source consisted of
an agile $15.5$~GHz source followed by an amplifier multiplier chain
with 150~mW output power at $124\,\mbox{GHz}$. The $15.5$~GHz
source is obtained by mixing, with a single-side-band mixer, the output
of a 15.2~GHz dielectric resonator oscillator (DRO) with the output
of a $300$~MHz direct digital synthesizer (DDS) that is under field
programmable gate array (FPGA) control. On/off switching of the $124$~GHz
microwaves is done at $15.5$~GHz before the amplifier multiplier
chain. The microwaves are transported to the ions down the bore of
the magnet with a rigid waveguide and directed onto the ions with
a horn located between the ring and endcap electrodes of the trap.
With this arrangement the microwave hardware does not block optical
access along the magnetic field axis, enabling imaging of the ion
resonance fluorescence scattered along the magnetic field (top-view
image --- see~\ref{sec:wavefrontAlignment}). We obtain $\pi$-pulses
of $70\,\mu$s duration with the 150~mW output power of the amplifier
multiplier chain and the setup described here. The measured spin-echo
coherence duration ($T_{2}$) is $\sim100$~ms. 

At the end of an experimental sequence we turn on the Doppler cooling
laser and make a projective measurement of the ion spin state through
state-dependent resonance fluorescence. With the Doppler cooling laser
on, an ion in the $\left|\uparrow\right\rangle $ state scatters $\sim10^{7}$
photons/s while an ion in $\left|\downarrow\right\rangle $ is dark.
For the spin precession measurements reported here we performed a
global spin-state detection. Specifically we detected, with $f/5$
light collection and a photomultiplier tube, the resonance fluorescence
from all the ions in a direction perpendicular to the magnetic field
(the side-view). For detection periods of $\sim50$~ms the detection
fidelity is high, typically limited by quantum projection noise. Here
we used short detection periods of $\sim500\,\mu\mbox{s}$, from which
we would detect $\sim1$ photon for each bright state $\left|\uparrow\right\rangle $.
Typically, each experiment was repeated $\sim100$ times and averaged,
resulting in a few percent uncertainty due to shot noise in the measurement
of $P(\uparrow)$.

The spin-precession signal used to benchmark spin-spin coupling in
the manuscript relied on a global spin-state measurement via side-view
fluorescence collected on a photo-multiplier tube. In the future,
we anticipate that time-resolved top-view images such as that shown
in Fig.~1 will be used to obtain the spin state of individual ions.
As discussed in the manuscript, ion crystal rotation at $\omega_{r}$
is phase-locked locked to an external oscillator. We use an imaging
photomultiplier tube to record $(x,y,t)$ for each photon. Rotating-frame
images are generated computationally given $(x,y,t)$ and $\omega_{r}$,
a technique established in 2001 \cite{Mitchell2001}. In linear Paul
trap experiments, determination of ion spin-state is possible with
as few as $10$ photons/ion \cite{wineland1998bible}. At present
we await the arrival of a new $(x,y,t)$ detector system capable of
a detection rate of $5\times10^{6}$~Hz. We anticipate this will
enable high-fidelity spin-state measurement of a $300$-ion crystal
in $\sim10$~ms. From a suite of identical experiments, the spin-spin
correlation function can be computed. We believe that our ability
to resolve single ions, even in the presence of rotation at $\omega_{R}$
indicates a path forward in performing individually resolved measurements
of fluorescence correlations between ions.

Previous measurements have elucidated some of the possible limiting
mechanisms. For instance, the stability of crystal orientation in
the rotating frame was studied for spherical crystals of $\sim10,000$
ions. In these experiments the orientation was observed to precess
uniformly for durations of $\gtrsim10$~s, then suddenly slip by
a large angle before again resuming slow precession \cite{Mitchell2001}.
This ``stick-slip'' motion can be followed and easily corrected.
Figure~1a of the manuscript was generated from $600$~s of integration
after stick-slip corrections. Other potential issues include ion loss
and background gas collisions. As the trap depth is $\gg1$~eV, background
gas collisions do not result ion loss. Collisions with background
hydrogen generate $\mbox{BeH}^{+}$ (1 per 6 minutes for $N\sim300$
$^{9}\mbox{Be}^{+}$ ions) which collect at the crystal perimeter
due to centrifugal separation. These effects do not appear to limit
top-view imaging fidelity, as demonstrated in Fig.~1a of the manuscript. 

\section{Optical-dipole-force laser settings}

\label{sec:odfLaserSettings}

\begin{figure}[t]
\begin{centering}
\includegraphics[width=0.95\columnwidth]{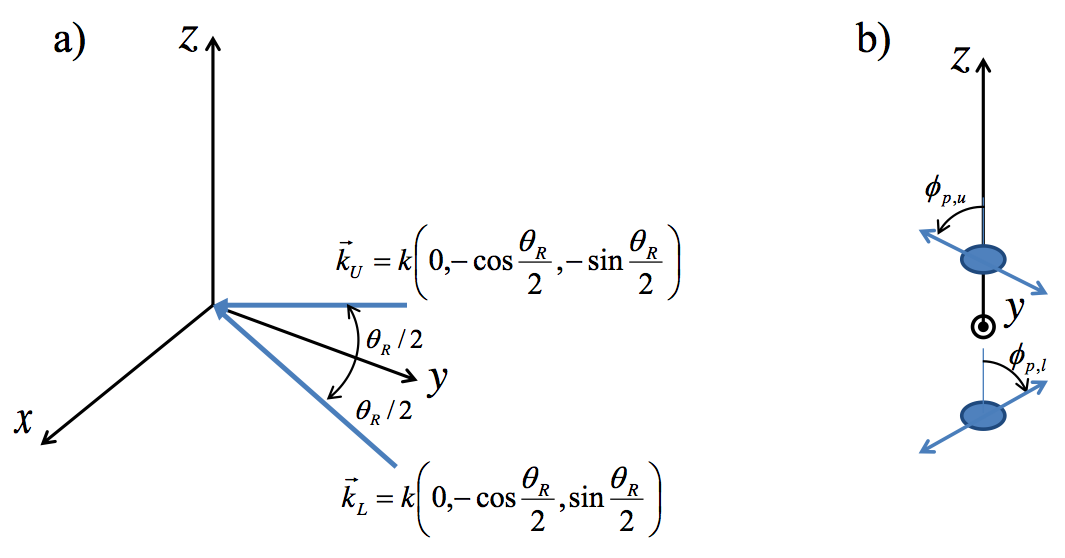}
\par\end{centering}

\caption{Sketch of optical dipole force (ODF) laser beam setup. (a) The ODF
laser beams lie in the $y$-$z$ plane at angles $\pm\theta_{R}/2$
with respect to the $y$-axis. (b) View looking in the $-\hat{y}$
direction. The beams are linearly polarized but with different polarization
angles relative to vertical polarization.\label{fig:ODF_setup}}
\end{figure}

Figure \ref{fig:ODF_setup} shows a simple sketch of the optical dipole
force (ODF) laser beam set-up. As discussed below, the frequency as
well as the beam polarizations were chosen to null the AC~Stark shift
from an individual beam and to produce a state-dependent force that
is equal in magnitude but opposite in sign for the $\left|\uparrow\right\rangle $
and $\left|\downarrow\right\rangle $ qubit states ($F_{\uparrow}=-F_{\downarrow}$
). This reduced the system sensitivity to laser intensity fluctuations.
For example, if $F_{\uparrow}\neq-F_{\downarrow}$, then the interaction
induced by the optical dipole force will include terms linear in the
$\hat{\sigma}_{i}^{z}$'s. These terms can be canceled with spin-echo
techniques, but this requires that laser intensity fluctuations are
small. Likewise, by adjusting the laser polarization to null the AC
Stark shift from a single beam, we mitigated qubit decoherence due
to laser intensity fluctuations. For the benchmarking measurements
described here we did not actively stabilize (i.e., noise eat) the
laser beam intensity.

The off-resonant laser beam frequency was detuned from the cycling
transition $\left(\left|\uparrow\right\rangle \rightarrow\left|P_{3/2},m_{J}=3/2\right\rangle \right)$
by $\Delta_{R}\simeq-63.8$~GHz. This gives detunings of $+15.6$~GHz
and $-26.1$~GHz, respectively, from the $\left|\uparrow\right\rangle \rightarrow\left|P_{3/2},m_{J}=1/2\right\rangle $
and $\left|\downarrow\right\rangle \rightarrow\left|P_{3/2},m_{J}=-1/2\right\rangle $
transitions (Fig.~\ref{figLevelDiagram}). Laser beam waists were
$w_{z}\simeq110\:\mu$m in the vertical (z-direction) and $w_{x}\simeq1$~mm
in the horizontal direction. Here we define the waist as the distance
from the center of the beam over which the electric field intensity
decreases by $1/e^{2}$ (i.e. $I(z)\sim e^{-(z/w_{z})^{2}}\:$). With
the small $2.4^{\circ}$ incident angle each beam makes with respect
to the plane of the crystal, this provided a uniform electric field
with $<10\%$ intensity variation across ion crystal arrays with $N<250$.

The ODF laser beams were linearly polarized at nonzero angles with
respect to the $\hat{z}$-axis. Let 
\begin{equation}
\begin{array}{ccc}
\vec{E}_{U}\left(\vec{r},t\right) & = & \hat{\epsilon}_{U}E_{U}\cos\left(\vec{k}_{U}\cdot\vec{r}-\omega_{U}t\right)\\
\vec{E}_{L}\left(\vec{r},t\right) & = & \hat{\epsilon}_{L}E_{L}\cos\left(\vec{k}_{L}\cdot\vec{r}-\omega_{L}t\right)
\end{array}
\end{equation}
denote the electric fields of the upper and lower ODF beams. If $\phi_{p}$
is the angle of the laser beam electric-field polarization with respect
to vertical polarization $\left(\hat{\epsilon}\cdot\hat{x}=0\right)$,
then the AC Stark shift of the qubit states when illuminated by a
single beam can be written
\begin{equation}
\begin{array}{c}
\Delta_{\uparrow,\, acss}=A_{\uparrow}\cos^{2}\left(\phi_{p}\right)+B_{\uparrow}\sin^{2}\left(\phi_{p}\right)\\
\Delta_{\downarrow,\, acss}=A_{\downarrow}\cos^{2}\left(\phi_{p}\right)+B_{\downarrow}\sin^{2}\left(\phi_{p}\right)
\end{array}
\end{equation}
where $A_{\uparrow}$($A_{\downarrow}$) is the Stark shift of the
$\left|\uparrow\right\rangle $($\left|\downarrow\right\rangle $)
state for a $\pi$-polarized beam ($\hat{\epsilon}$ parallel to the
$\hat{z}$-axis) and $B_{\uparrow}$($B_{\downarrow}$) is the Stark
shift of the $\left|\uparrow\right\rangle $($\left|\downarrow\right\rangle $)
state for a $\sigma$-polarized beam ($\hat{\epsilon}$ perpendicular
to the $\hat{z}$-axis). (Here we neglect the small $\sigma$ polarization
($\propto\sin2.4^{o}$) that exists when $\phi_{p}=0$.) The Stark
shift of the qubit transition is 
\begin{equation}
\Delta_{acss}=\left(A_{\uparrow}-A_{\downarrow}\right)\cos^{2}\left(\phi_{p}\right)+\left(B_{\uparrow}-B_{\downarrow}\right)\sin^{2}\left(\phi_{p}\right)\:.\label{eq:ACStark_shift}
\end{equation}
If $A_{\uparrow}-A_{\downarrow}$ and $B_{\uparrow}-B_{\downarrow}$
have opposite signs, there is an angle which makes $\Delta_{acss}=0$.
For a laser detuning of $\Delta_{R}=-63.8$ GHz, $\Delta_{acss}=0$
at $\phi_{p}\simeq\pm65{}^{o}$.

With $\Delta_{acss}=0$ for each ODF laser beam, we exploit the freedom
to choose their polarization in order to obtain a state-dependent
force. Specifically, we choose $\vec{E}_{U}$ to have a polarization
given by $\phi_{p,u}=65{}^{o}$ and $\vec{E}_{L}$ to have a polarization
given by $\phi_{p,l}=-65{}^{o}$. In this case the interference term
in the expression for the electric field intensity $\left(\vec{E}_{U}+\vec{E}_{L}\right)^{2}$
produces a polarization gradient which results in spatially dependent
AC Stark shifts 
\begin{equation}
\begin{array}{c}
\left(A_{\uparrow}\cos^{2}\left(\phi_{p}\right)-B_{\uparrow}\sin^{2}\left(\phi_{p}\right)\right)2\cos\left(\delta k\cdot z-\mu_{R}t\right)\\
\left(A_{\downarrow}\cos^{2}\left(\phi_{p}\right)-B_{\downarrow}\sin^{2}\left(\phi_{p}\right)\right)2\cos\left(\delta k\cdot z-\mu_{R}t\right)
\end{array}
\end{equation}
for the qubit levels. Here $\delta k\equiv\left|\vec{k}_{U}-\vec{k}_{L}\right|=2k\sin\left(\frac{\theta_{R}}{2}\right)$
is the wave vector difference between the two ODF laser beams, $\mu_{R}=\omega_{U}-\omega_{L}$
is the ODF beat note, and $\phi_{p}=\left|\phi_{p,u}\right|=\left|\phi_{p,l}\right|$.
The spatially dependent AC Stark shift produces a state-dependent
force $F_{\uparrow,\downarrow}(z,t)=F_{o\:\uparrow,\downarrow}\sin\left(\delta k\cdot z-\mu_{R}t\right)$
where 
\begin{equation}
\begin{array}{c}
F_{o\uparrow}=-2\,\delta k\left(A_{\uparrow}\cos^{2}\left(\phi_{p}\right)-B_{\uparrow}\sin^{2}\left(\phi_{p}\right)\right)\\
F_{o\downarrow}=-2\,\delta k\left(A_{\downarrow}\cos^{2}\left(\phi_{p}\right)-B_{\downarrow}\sin^{2}\left(\phi_{p}\right)\right).
\end{array}
\end{equation}
In general $F_{o\uparrow}\neq-F_{o\downarrow}.$ We operate at $\Delta_{R}=-63.8$~GHz
where for $\Delta_{acss}=0$ we also obtain $F_{o\uparrow}=-F_{o\downarrow}\equiv F_{o}$ 

For a given $\phi_{p,u}\,$, $\phi_{p,l}\,$, and $\Delta_{R}$ we
use straightforward atomic physics along with well known values for
the energy levels and matrix elements of $\nbep$ to calculate $F_{o}$
as a function of the electric field intensity $I_{R}=\frac{c\epsilon_{o}}{2}\left|E_{L}\right|^{2}=\frac{c\epsilon_{o}}{2}\left|E_{U}\right|^{2}$
at the center of the laser beams. For $\theta_{R}=4.8^{\circ}$ and
$I_{R}=1\:$~W/cm$^{2}$ , $F_{o}=1.4\times10^{-23}$~N.

Stronger forces can be generated after experimental modification to
our apparatus to permit $\theta_{R}\approx35^{o}$ . At this angle
larger detunings $\mu_{R}-\omega_{1}$ are required to satisfy Eq.~6
in Methods. With our definition $H_{I}=\frac{1}{N}\sum_{i<j}J_{i,j}\hat{\sigma}_{i}^{z}\hat{\sigma}_{j}^{z}$,
the interaction strength between two ions $i$ and $j$ is $J_{i,j}/N$.
Consider for example $N=217$~ions, $\omega_{r}=2\pi\times45.6$~kHz,
powers of $I_{R}=20$~mW/beam ($12.5\,\mbox{W}/\mbox{cm}^{2}$) and
$\theta_{R}=35^{o}$. The spin-motion entanglement constraint is satisfied
by a detuning of $\mu_{R}-\omega_{1}=2\pi\times100$~kHz. In this
case we obtain $J_{i,j}/(2\pi N)\sim\left(560\: Hz\right)\left(d_{0}/d_{i,j}\right)^{1.7}$,
where $d_{0}\sim20\,\mbox{\ensuremath{\mu}m}$ is the typical nearest
neighbor separation.

\section{Wavefront alignment}

\label{sec:wavefrontAlignment}

\begin{figure}
\begin{centering}
\includegraphics[width=1\columnwidth]{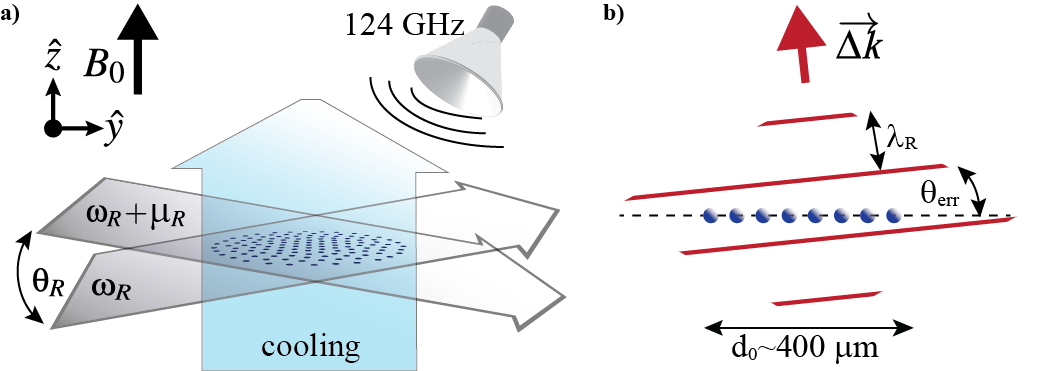}
\par\end{centering}

\caption{\label{fig:tilted_wavefronts}(a) Sketch of the ODF laser beam geometry
used to generate the one dimensional (1D) traveling optical lattice.
(b) Sketch of the 1D optical lattice wave fronts (red lines). These
wave fronts need to be aligned with with the ion planar array (represented
by the blue dots). Here $\lambda_{R}=2\pi/|\protect\overrightarrow{\Delta k}|\approx3.7\,\mu\mbox{m}$
and $\theta_{err}$ denotes the angle of misalignment. $400\:\mu$m
is the typical array diameter for $N\sim200$ ions. With the wavefront
alignment technique discussed in the text we obtain $\theta_{\mbox{err}}<0.05^{\circ}$.}
\end{figure}

\begin{figure}[b]
\begin{centering}
\includegraphics[width=0.8\columnwidth]{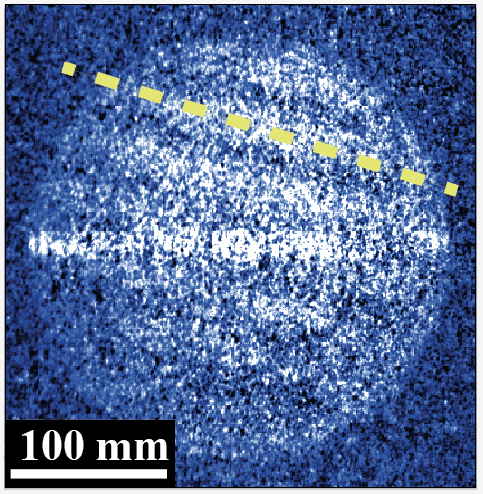}
\par\end{centering}

\caption{\label{fig:wavefront_align}Top-view image of the spatially inhomogeneous
fluorescence from a single ion plane produced by the AC~Stark shift
from a static $(\mu_{R}=0)$ ODF lattice with misaligned wave fronts.
Dark bands are regions of high standing wave electric field intensity
(parallel to the dashed yellow line). The bright horizontal feature
bisecting the center of the image is fluorescence from the weak Doppler
laser cooling beam directed perpendicular to the magnetic field. The
image was obtained by subtracting a background image with the ODF
beams off ($1$~s integration).}
\end{figure}
The ODF laser beams produce a one dimensional (1D) optical lattice
characterized by the effective wave vector $\delta\vec{k}$ and beat
note $\mu_{R}$. In Sec.~\ref{sec:odfLaserSettings} we assumed that
$\delta\vec{k}\parallel\hat{z}$, or equivalently that the wavefronts
of the lattice were aligned perpendicular to the $\hat{z}$-axis (magnetic
field axis). If the wavefronts are not normal to the $\hat{z}$-axis
as sketched in Fig.~\ref{fig:tilted_wavefronts}, then the time dependence
of the ODF seen by an ion in the rotating frame depends on the $(x,y)$
position of the ion. This complicates the effective spin-spin interactions
generated by the ODF but can be adequately mitigated by careful alignment.

Alignment of the ODF laser beams is obtained by a technique that makes
top-view images (images of the ion resonance fluorescence scattered
along the magnetic field) from a single plane of ions sensitive to
ODF wave front misalignment. For this measurement we set $\mu_{R}=0$
(stationary 1D lattice) and detune the frequency of the ODF laser
beams approximately $0.5$~GHz below the $\left|\uparrow\right\rangle \rightarrow\left|^{2}P_{3/2}\: m_{J}=+3/2\right\rangle $
Doppler cooling transition. This small detuning generates sufficiently
large AC~Stark shifts on the cooling transition to measurably change
the ion scatter rate from the Doppler cooling laser. With the Doppler
cooling laser turned on and the ODF beams turned off, we observe a
spatially uniform, time-averaged image of a rotating planar crystal.
With the ODF beams on, ions located in regions of high electric field
intensity at the anti-nodes of the optical lattice are Stark-shifted
out of resonance with the Doppler cooling laser. This is the cause
of the dark bands in the top-view image shown in Fig.~\ref{fig:wavefront_align}.
We adjust the ODF laser beams based on this real-time imaging to optimize
their alignment. Improved alignment is indicated by a fringe pattern
of longer wavelength. With this technique we have aligned the ODF
wave fronts with the planar array to better than $\sim0.05^{\circ}$. 

The image in Fig.~\ref{fig:wavefront_align} is typical of what we
obtain with $1\,\mbox{s}$ integration duration. This indicates the
1D lattice was stable during the integration period and shows the
phase stability of our 1D lattice of better than $1\,\mbox{s}$. 

We note that direct fluorescence imaging of the 1D~lattice, for example
by tuning the ODF laser resonant with the Doppler cooling transition,
is not viable. Even at low powers, resonantly scattered photons across
the large horizontal waist of the ODF beams apply a large torque,
causing the rotation frequency and radius of the array to rapidly
change, typically driving the ions into orbits of very large radius. 

We have also used phase coherent Doppler velocimetry to improve the
ODF wavefront alignment \cite{Biercuk2011}. But the top-view imaging
technique discussed here and shown in Fig.~\ref{fig:wavefront_align}
provides more information on the angle and direction of misalignment,
which greatly improves the ODF-crystal alignment process.

\section{Modeling mean field spin precession}

\label{sec:precessionFormulaDerivation}

\begin{figure}
\includegraphics[width=0.9\columnwidth]{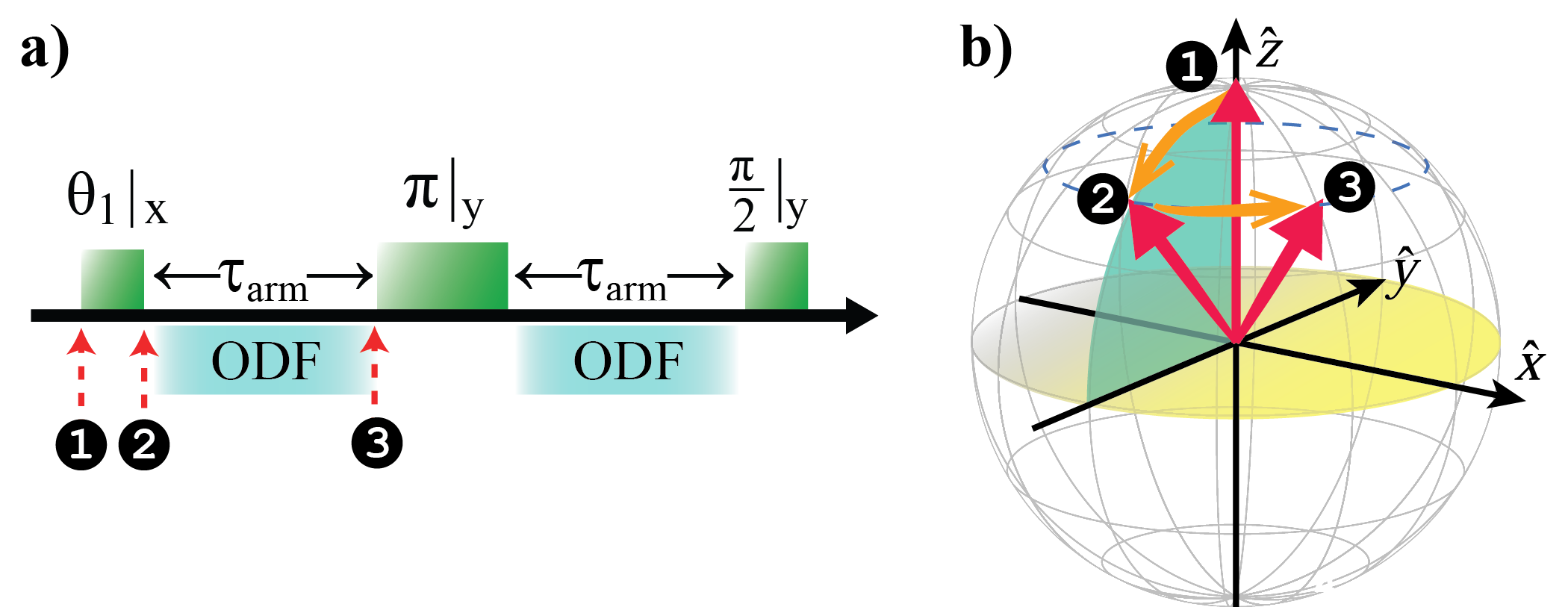}

\caption{(a) Sequence used to measure spin precession due to the effective
mean field generated by the engineered Ising interaction. The spins
are initially prepared in $\left|\uparrow\right\rangle $. The first
pulse rotates the spins by an angle $\theta_{1}$ about an axis (defined
to be the $\hat{x}$-axis in the rotating frame) in the equatorial
plane. The subsequent $\pi$ and $\pi/2$ pulses occur about the $\hat{y}$-axis.
(b) Evolution of a single spin prior to the application of the spin-echo
$\pi$-pulse.}

\label{fig:benchmark_fig}
\end{figure}

For any spin $j$, $\left[\hat{\sigma}_{j}^{z},\hat{H}_{I}\right]=0$,
which implies $\left\langle \hat{\sigma}_{j}^{z}\right\rangle =$
constant under application of $\hat{H}_{I}$. The Hamiltonian $\hat{H}_{I}^{MF}$
describing the mean field response of spin $j$ to the application
$\hat{H}_{I}$ is obtained by expanding $\hat{H}_{I}$ to first order
in $\widehat{\delta\sigma}_{j}^{z}\equiv\hat{\sigma}_{j}^{z}-\left\langle \hat{\sigma}_{j}^{z}\right\rangle $:
\begin{equation}
\begin{array}{ccc}
\hat{H}_{I}^{MF} & = & \frac{1}{N}\sum_{i<j}J_{i,j}\left(\left\langle \hat{\sigma}_{i}^{z}\right\rangle \widehat{\delta\sigma}_{j}^{z}+\left\langle \hat{\sigma}_{j}^{z}\right\rangle \widehat{\delta\sigma}_{i}^{z}\right)\\
 & = & \sum_{j=1}^{N}\frac{1}{N}\sum\limits _{i,i\neq j}^{N}J_{i,j}\left\langle \hat{\sigma}_{i}^{z}\right\rangle \times\widehat{\delta\sigma}_{j}^{z}.
\end{array}
\end{equation}
Defining $\bar{B}_{j}\equiv\frac{2}{N}\sum_{i,i\neq j}^{N}J_{i,j}\left<\hat{\sigma}_{i}^{z}\right>$,
$\hat{H}_{I}^{MF}$ can be written
\begin{equation}
\hat{H}_{I}^{MF}=\sum_{j=1}^{N}\bar{B}_{j}\widehat{\delta\sigma}_{i}^{z}/2.\label{eq:HIMF}
\end{equation}
The mean field Heisenberg equations of motion for spin $\vec{\sigma}_{k}$
is obtained from the commutator of $\vec{\sigma}_{k}$ and $\hat{H}_{I}^{MF}$.
However, because $\widehat{\delta\sigma}_{j}^{z}=\hat{\sigma}_{k}^{z}-\left<\hat{\sigma}_{k}^{z}\right>$
and $\left<\hat{\sigma}_{k}^{z}\right>$ is a constant, it is clear
that Eq.~\ref{eq:HIMF} describes spin precession about the z-axis
at frequency $\bar{B}_{k}$. Our observable is the spin precession
(in excess of ordinary Larmor precession) averaged over all the spins
$\frac{1}{N}\sum_{k=1}^{N}\bar{B}_{k}$. 

We use the spin-echo sequence in Fig.~\ref{fig:benchmark_fig} to
measure a precession proportional to the expectation value of the
spin projection along $\hat{z}$ $\left(\left\langle \hat{\sigma}_{i}^{z}\right\rangle \right)$.
The spin-echo sequence minimizes contributions to spin precession
that are not $\propto$$\left\langle \hat{\sigma}_{i}^{z}\right\rangle $.
Specifically, the spin echo cancels a constant spin precession independent
of $\left\langle \hat{\sigma}_{i}^{z}\right\rangle $ (e.g., due to
slow uncontrolled magnetic field fluctuations), but precession $\propto$$\left\langle \hat{\sigma}_{i}^{z}\right\rangle $
coherently adds in the two arms of the sequence. 

The first pulse sets $\left\langle \hat{\sigma}_{i}^{z}\right\rangle =\cos\theta_{1}$.
The interaction $\hat{H}_{I}$ is then applied by turning on the ODF
laser beams for a period $\tau_{arm}$. During this interval $\left\langle \hat{\sigma}_{i}^{z}\right\rangle $
is a constant and mean field theory predicts an average spin precession
angle of $2\bar{J}\cos(\theta_{1})\cdot\tau_{arm}$. The $\pi$-pulse
changes $\left\langle \hat{\sigma}_{i}^{z}\right\rangle \rightarrow-\cos\left(\theta_{1}\right)$
and the spin precession angle $2\bar{J}\cos(\theta_{1})\cdot\tau_{arm}$
to $-2\bar{J}\cos(\theta_{1})\cdot\tau_{arm}$. At the end of the
second $\hat{H}_{I}$ interaction of duration $\tau_{arm}$, the total
precession angle of the spins is $-2\bar{J}\cos(\theta_{1})\cdot2\tau_{arm}$.
The final $\pi/2$-pulse is about an axis shifted by $90^{o}$ from
the first $\theta_{1}$-pulse. This pulse converts precession out
of the initial $\widehat{yz}$ plane into excursions above or below
the equatorial plane of the Bloch sphere, which we measure. 

The evolution operator of the measurement sequence $\hat{U}_{seq}$
is obtained in a straight forward manner from the individual evolution
operators from each segment of the sequence,
\begin{equation}
\hat{U}_{seq}=\hat{R}\left(\hat{y},\frac{\pi}{2}\right)\cdot\hat{U}(\hat{H}_{I}^{MF})\cdot\hat{R}\left(\hat{y},\pi\right)\cdot\hat{U}\left(H_{I}^{MF}\right)\cdot\hat{R}\left(\hat{x},\theta_{1}\right)\:.
\end{equation}
Here $\hat{R}\left(\hat{x},\theta_{1}\right)=\left[\begin{array}{cc}
\cos(\theta_{1}/2) & -i\sin(\theta_{1}/2)\\
-i\sin\left(\theta_{1}/2\right) & \cos\left(\theta_{1}/2\right)
\end{array}\right]$, $\hat{R}\left(\hat{y},\pi\right)=\left[\begin{array}{cc}
0 & -1\\
1 & 0
\end{array}\right]$ and $\hat{R}\left(\hat{y},\frac{\pi}{2}\right)=\frac{\sqrt{2}}{2}\left[\begin{array}{cc}
1 & -1\\
1 & 1
\end{array}\right]$. The mean field evolution is given by 
\begin{equation}
\hat{U}\left(\hat{H}_{I}^{MF}\right)=\left[\begin{array}{cc}
\exp\left(-i\bar{J}\left\langle \hat{\sigma}_{z}\right\rangle \tau_{arm}\right) & 0\\
0 & \exp\left(i\bar{J}\left\langle \hat{\sigma}_{z}\right\rangle \tau_{arm}\right)
\end{array}\right]
\end{equation}
where $\left\langle \hat{\sigma}_{z}\right\rangle =\cos\left(\theta_{1}\right)$
in the first arm of the spin-echo sequence and $\left\langle \hat{\sigma}_{z}\right\rangle =-\cos\left(\theta_{1}\right)$
in the second arm. At the end of the sequence we detect the $\left|\uparrow\right\rangle $
state probability. Explicit computation gives 
\begin{equation}
\left|\left<\uparrow|\hat{U}_{seq}|\uparrow\right>\right|^{2}=\frac{1}{2}\left\{ 1+\sin\left(\theta_{1}\right)\cdot\sin\left[2\bar{J}\cos\left(\theta_{1}\right)\cdot2\tau_{arm}\right]\right\} \:.\label{eq:benchmark_no_Gamma}
\end{equation}

\section{Optical dipole force laser intensity calibration}

\label{sec:odfLaserIntensityCal}

\begin{figure}
\begin{centering}
\includegraphics[width=0.8\columnwidth]{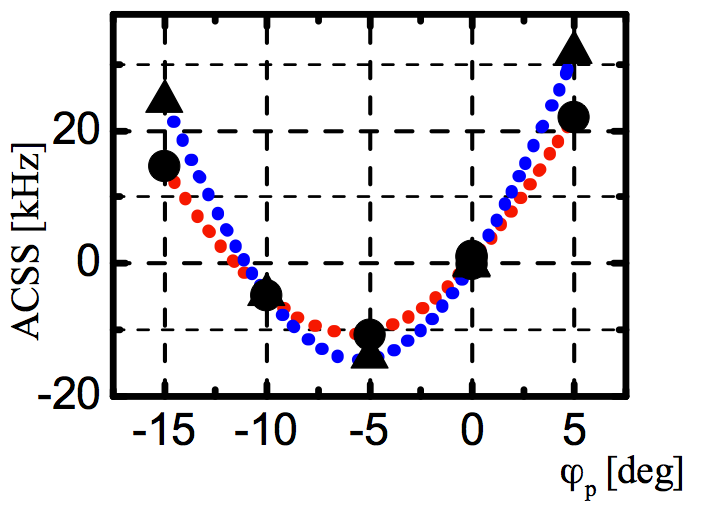} 
\par\end{centering}

\caption{We determine $I_{R}$ from the AC Stark shift of the qubit transition
as a function of ODF laser beam polarization angle $\phi_{p}$. The
triangles and filled circles are separate measurements done on the
upper and lower ODF beams. $I_{R}$ is determined independently for
each ODF beam by a fit to Eq.~\ref{eq:ACStark_shift}. The error
bars are smaller than the plot points.}

\label{fig:SM_acss}
\end{figure}

The spin-spin couplings ($\bar{J}$) measured here depend on the square
of the ODF electric field intensity $I_{R}$. Therefore, careful calibration
of the electric field intensity was important in benchmarking the
strength of the interactions with mean-field predictions. We separately
determined $I_{R}$ for each beam by measuring the AC Stark shift
of the qubit transition as a function of the laser beam polarization
angle $\phi_{p}$, as shown in Fig. \ref{fig:SM_acss}. The qubit
transition frequency was measured by fitting for the center frequency
of a Rabi-resonance profile. The polarization angle $\phi_{p}$ was
varied by rotating a $\lambda$/2 plate. The Stark shift measurements
were fit to Eq. \ref{eq:ACStark_shift}, which provided values for
$A$ and $B$ (Stark shifts for $\pi$-polarization and $\sigma$-polarization).
Values for $A$ and $B$ were then used with atomic physics calculations
to determine $I_{R}$, the electric field intensity of the ODF laser
beams, with a fractional uncertainty of $\sim5\%$. Frequent intensity
calibration measurements were taken during a benchmarking run. Slow
drifts to the laser intensity between calibrations added another $5\%$
uncertainty, which we add in quadrature to the fitted uncertainty.

\section{Spontaneous emission}

\label{sec:odfSpontEm}

Decoherence due to spontaneous emission has been well studied in this
system \cite{Uys2010}. The qubit levels are closed under spontaneous
light scattering; that is, spontaneous light scattering does not optically
pump the ion to a different ground state level outside of the two
qubit levels. We measure spin precession to benchmark the Ising interaction
couplings. Equivalently we measure the evolution of off-diagonal coherences
between the $\left|\uparrow\right\rangle $ and $\left|\downarrow\right\rangle $
levels. Spontaneous-emission-induced decay of these coherences is
accurately modeled by Eq.~(8) of Ref.~\cite{Uys2010}. We add this
time dependence to the unitary evolution $\hat{U}\left(\hat{H}_{I}^{MF}\right)$
discussed in Sec.~\ref{sec:precessionFormulaDerivation}. This results
in the expression 
\begin{equation}
P\left(\left|\uparrow\right\rangle \right)=\frac{1}{2}\left(1+\exp\left(-\Gamma\cdot2\tau_{arm}\right)\sin\left(\theta_{1}\right)\sin\left(2\bar{J}\cos\left(\theta_{1}\right)\cdot2\tau_{arm}\right)\right)\,,\label{eq:benchmarking}
\end{equation}
where $P\left(\left|\uparrow\right\rangle \right)$ is the probability
of detecting a spin in $\left|\uparrow\right\rangle $ at the end
of the spin-echo sequence. The only difference with Eq.~\ref{eq:benchmark_no_Gamma}
is the factor $\exp(-\Gamma\cdot2\tau_{arm})$ due to the decay of
the off-diagonal elements of the density matrix. Here $\Gamma$ accounts
for decoherence due to spontaneous emission. From \cite{Uys2010},
$\Gamma=\frac{1}{2}\left(\Gamma_{Ram}+\Gamma_{el}\right)$ has contributions
from both Raman scattering and elastic Rayleigh scattering. A straightforward
atomic physics calculation relates $\Gamma$ to the ODF laser beam
intensity $I_{R}$. In fits of Eq.~\ref{eq:benchmarking} to the
spin-precession measurements, we fix $\Gamma$ at the value determined
by the laser intensity calibrations. For $\Delta_{R}=-63.8$ GHz,
$\phi_{p}=\pm65.3^{o}$, and $I_{R}=1$ W/cm$^{2}$, we calculate
$\Gamma=82$ s$^{-1}$. 

With our present set-up we can achieve $\bar{J}\gg\Gamma$ for detunings
$\left|\mu_{R}-\omega_{1}\right|\lesssim10$ kHz. For these detunings
it will be possible to simulate quantum effects beyond mean field
theory such as spin squeezing and spin-depolarization due to many-body
interactions. For example we calculate that with $4$~mW/beam and
an ODF beatnote detuning of $\left|\mu_{R}-\omega_{1}\right|=2$~kHz
, $\bar{J}\sim2\pi\times0.5$~kHz and $\Gamma/\bar{J}\sim0.06$.
The potential spin squeezing due to this interaction is $5$~dB,
limited by spontaneous emission. For many simulations it will be desirable
to achieve $J/N\gg\Gamma$ where $J$ is a nearest-neighbor coupling
strength. ($J\sim\bar{J}$ for small detunings $\left|\mu_{R}-\omega_{1}\right|$.)
The most straight forward strategy to achieve this condition in our
set-up appears to be to increase the angle $\theta_{R}$ between the
Raman beams. The ratio $\bar{J}/\Gamma$ scales as $\sin^{2}\left(\theta_{R}/2\right)$.
Therefore decoherence due to spontaneous emission can be dramatically
reduced with an order of magnitude increase in $\theta_{R}.$ We note
that a large increase in $\theta_{R}$ will require implementation
of a sub-Doppler cooling scheme to remain in the Lamb-Dicke regime
(see Sec.~\ref{sec:Lamb-DickeConfinement}). 

Different ODF laser detunings can likely help reduce the impact of
spontaneous emission (with the complication that $F_{\uparrow}\neq-F_{\downarrow}$),
but very large laser detunings $\Delta$ obtained by tuning the ODF
laser beam frequencies outside the $P_{1/2}-P_{3/2}$ manifold appear
unlikely to help. This is because both the interaction strength and
spontaneous emission scale as $1/\Delta^{2}$ for our qubit, which
is not a clock transition. For trapped-ion experiments in low magnetic
field, the impact of spontaneous emission is typically minimized by
tuning the ODF laser beam frequencies between the $P_{1/2}$ and $P_{3/2}$
manifolds. Because the $^{9}$Be$^{+}$ $P$-state fine structure
($197$~GHz) is comparable to the Zeeman splittings for our magnetic
field ($4.5$~T), this approach does not significantly help. However,
it can help with heavier mass ions (e.g.,~Mg$^{+}$). Reference~\cite{Baltrusch2011}
discusses a number of strategies that can minimize the impact of spontaneous
emission when the $P$-state fine structure is large compared to the
Zeeman interaction energy.

\section{Lamb-Dicke confinement}

\label{sec:Lamb-DickeConfinement}

Throughout this manuscript we implicitly assume that the applied state-dependent
force, $F_{\uparrow,\downarrow}(z,t)=F_{o\:\uparrow,\downarrow}\cos\left(\delta k\cdot z-\mu_{R}t\right)$,
is constant across the spatial extent of an ion's wave function. Specifically,
for a planar array located at $z=0$, we assume the force on any ion
is given by 
\begin{equation}
F_{\uparrow,\downarrow}\left(z=0,t\right)=F_{o\uparrow,\downarrow}\cos\left(\mu_{R}t\right)\:.\label{eq:forceLambDicke}
\end{equation}
The extent to which this is true is quantified by the parameter
\begin{equation}
\eta_{ind,i}\equiv\delta k\cdot z_{rms,i}
\end{equation}
where $z_{rms,i}$ is the root mean square (rms) axial extent of the
wave function of ion $i$ and $\delta k\equiv2\pi/\lambda_{R}$, where
$\lambda_{R}=3.7\:\mu$m for $\theta_{R}=4.8^{o}$. We note that $\eta_{ind,i}$
is a Lamb-Dicke \emph{confinement} parameter for a individual ion
(ind), not to be confused with the usual Lamb-Dicke parameter that
is defined in terms of the ground-state wave function. In the limit
that the Coulomb interaction energy between ions is a small perturbation
to the axial potential of the trap, we can think of each ion as a
single ion confined in the external trap potential, and approximate
$z_{rms,i}\simeq\sqrt{\frac{\hslash}{2M\omega_{z}}}\cdot\sqrt{2\bar{n}+1}$,
where $M$ is the mass of an individual ion. Recently we have completed
careful axial temperature measurements for single-plane ion crystals
\cite{Sawyer2012modeAndTempSpec}. We find $\sim1$~mK for the axial
COM mode and $\sim0.4$~mK for the higher order transverse modes.
For $\omega_{z}=2\pi\cdot795$ kHz and conservatively assuming $T=1$~mK,
we calculate $\bar{n}\simeq26$, $z_{rms,i}\simeq190$~nm, and $\eta_{ind,i}\simeq0.32$. 

For the benchmarking measurements described here we typically set
the rotation frequency of the ion array about $\sim0.5$~kHz below
the rotation frequency of the $1\leftrightarrow2$ plane transition,
determined experimentally from side-view images \cite{Mitchell1998}.
In this case, the spectrum of the transverse (axial) modes is broad,
and we underestimate $z_{rms,i}$ in the above analysis. An improved
estimate of $z_{rms,i}$ is obtained by summing the contributions
from all of the transverse modes $m$
\begin{equation}
z_{rms,i}=\left(\sum_{m}\left(b_{i,m}\right)^{2}\frac{\hbar}{2M\omega_{m}}\left(2\bar{n}_{m}+1\right)\right)^{1/2},\label{eq:zrms}
\end{equation}
where $\bar{n}_{m}\simeq k_{B}T/\hbar\omega_{m}$ is the mean thermal
occupation of mode $m$. We assume here that every mode is characterized
by the same temperature T. For $N=217$ and $\omega_{z}=2\pi\times795$~kHz,
the rotation frequency of the $1\leftrightarrow2$ plane transition
is $\omega_{r}=2\pi\times46.1$~kHz. For $\omega_{R}=2\pi\times45.6$~kHz,
the spectrum of the axial modes ranges from $795$~kHz down to $224$~kHz.
With Eq.~\ref{eq:zrms} we calculate $z_{rms,i}\simeq520$~nm in
the center of the array, decreasing to $z_{rms,i}\simeq250$~nm at
the array edge, corresponding to $\eta_{ind,i}\simeq0.89$ for $i$
near the array center and $\eta_{ind,i}=0.42$ for $i$ near the array
boundary.

The wavefront alignment discussed in a previous section can be viewed
as achieving a type of Lamb-Dicke confinement. Let $\theta_{err}$
denote the angle of misalignment between the planar array and the
ODF 1D lattice (Fig.~\ref{fig:tilted_wavefronts}). Relative to the
ODF lattice wavefronts, the rotation of the array produces a time-dependent
shift in the axial position of an ion that can be written as $z(t)=R\sin\left(\theta_{err}\right)\sin\left(\omega_{r}t+\varphi\right)$.
Here R is the distance of the ion from the center of the array, and
$\varphi$ is determined by the azimuthal position of the ion in the
array. The force on this ion is then 
\begin{equation}
F_{\uparrow,\downarrow}\left(t\right)=F_{o\uparrow,\downarrow}\cos\left(\delta kR\sin\theta_{err}\cdot\sin\left(\omega_{r}t+\varphi\right)-\mu_{R}t\right)\:.\label{eq:tilt_Force}
\end{equation}
For Eq. \ref{eq:tilt_Force} to approximate Eq. \ref{eq:forceLambDicke},
we desire $\delta k\cdot2R_{P}\sin\theta_{err}<1$, where $R_{P}$
is the array radius. For $R_{p}\simeq200\:\mu$m (typical for $N=200$)
and $\theta_{err}\simeq0.05^{\circ}$ (see Wavefront alignment section),
we calculate $\delta k\cdot2R_{P}\sin\theta_{err}\approx0.6$.

\section{Ion lattice configuration at equilibrium and transverse normal modes}

\label{sec:transModeCalc}

The ion equilibrium positions $\vec{r}_{i}$ at zero temperature are
calculated by minimizing the Euler-Lagrange action for the Penning
trap potentials and the constraint that the ions lie in a plane (at
$z=0$). The solution is a triangular lattice with a lattice constant
that increases as one moves radially outward, that has a smooth unfaceted
edge and exhibits a degradation in orientational order near the crystal
perimeter. The transverse (along $\hat{z}$) phonon modes ($\omega_{m}$,
$\vec{b}_{m}$) are obtained by Taylor expansion of the potential
about the ion equilibrium positions $\vec{r}_{i}$. In-plane modes
(along $\hat{x},\hat{y}$) can also be calculated, by solving a quadratic
eigenvalue problem (due to inclusion of the centrifugal and Coriolis
forces). The details of the transverse mode calculation is discussed
in this section. The problem has also been solved for longitudinal
and transverse modes in 1D (e.g., \cite{James1998,Marquet2003}).

\begin{flushleft}
In general, the Lagrangian for a collection of $N$ ions with charge
$q$ and mass $M$ in an electromagnetic field $\phi_{i}-\vec{A}(\vec{r}_{i})$
is 
\begin{equation}
L=T-V=\sum_{i=1}^{N}\left\{ \frac{1}{2}M\dot{\vec{r}}_{i}^{2}-q\left(\phi_{i}-\vec{A}(\vec{r}_{i})\cdot\dot{\vec{r}_{i}}\right)\right\} ,
\end{equation}
where $ $$\vec{r}_{i}=(r_{i},\theta_{i},z_{i})$ is the coordinate
of ion $i$. In a Penning trap, the field consists of a uniform magnetic
field in the $\hat{z}$-direction ($\vec{B}=B_{0}\hat{z}$) and a
harmonic trapping (anti-trapping) electric potential in the $\hat{z}$-direction
($\hat{r}$-direction) with frequency $\omega_{1}$. Lastly, an additional
external time-dependent electric quadrupole potential (the ``rotating
wall'', amplitude $V_{wall}$ at the ions) is applied to control
the ion's rotation frequency in the trap \cite{Bollinger2000}. Thus,
the scalar potential for ion $i$ is 
\begin{eqnarray}
\begin{array}{ccc}
q\phi_{i} & = & q\phi_{trap,i}+q\phi_{wall,i}+q\phi_{Coulomb,i}\\
 & = & \frac{1}{2}M\omega_{1}^{2}(z_{i}^{2}-r_{i}^{2}/2)+V_{wall}r_{i}^{2}\cos2(\theta_{i}+\omega_{r}t)+\frac{1}{2}k\sum_{j\neq k}^{N}q^{2}/|\vec{r}_{j,k}|,
\end{array}
\end{eqnarray}
where the ion-ion separation is $\vec{r}_{j,k}=\vec{r}_{j}-\vec{r}_{k}$
and $k=1/4\pi\epsilon_{0}$. Since $\vec{B}$ is uniform, the vector
potential energy is
\begin{equation}
q\vec{A}(\vec{r}_{i})=-\frac{q}{2}\vec{r_{i}}\times\vec{B}.
\end{equation}

\par\end{flushleft}

The position of the ions at equilibrium is calculated as follows.
We move to a rotating frame where the ions' coordinates are stationary
by using the coordinate transformation $\vec{r}_{i}'=(r_{i}',\theta_{i}',z_{i}')=(r_{i},\theta+\omega_{r}t,z_{i})$,
a counterclockwise rotation. The ions' equilibrium positions $\vec{r}_{i}'$
can be found by solving the transformed Euler-Lagrange equations $L'$.
To seed the numerical solution, we supply an initial guess for the
2D crystal: a regular, triangular lattice. Consistent convergence
requires ion numbers $N$ corresponding to closed shells (e.g., $N=127$
has six closed shells). We find ion equilibrium positions that deviate
from a perfect triangular lattice near the crystal periphery and which
have an overall ellipticity due to $V_{wall}$. 

Given $\vec{r}_{i}'$, the crystal's transverse eigenmodes can be
calculated by Taylor expansion of the potential about the equilibrium
positions $\vec{r}_{i}'$. The Lagrangian is 
\begin{equation}
L'=\frac{1}{2}\sum_{i=1}^{N}M\dot{z}_{i}'\dot{z}_{j}'-\frac{1}{2}\sum_{i,j=1}^{N}K_{ij}z_{i}'z_{j}'
\end{equation}
where $z_{i}'$ is the axial displacement of $i$-th ion and $K_{ij}=K_{ji}$
is the symmetric stiffness matrix evaluated for the equilibrium configuration
as 
\begin{equation}
K_{ij}=\left\{ \begin{array}{lll}
{\displaystyle {M\omega_{1}^{2}-\sum_{n=1}^{N}\frac{kq^{2}}{{|z_{ni}'|}^{3}}}} & \ \ \ \ i=j,n\neq i\\
{\displaystyle {\frac{kq^{2}}{{|\vec{r}_{ij}'|}^{3}}}} & \ \ \ \ i\neq j
\end{array}\right..
\end{equation}
By minimizing the action $\delta\int dtL'=0$ with respect to the
rotating frame axial coordinates $z_{i}'$, we obtain $N$ equations
of motion 
\begin{equation}
\ddot{z}_{i}'+\sum_{j=1}^{N}\frac{K_{ij}}{M}z_{j}'=0,\,\mbox{for }i=1,2,...,N.
\end{equation}
Following standard normal mode analysis, the solution is obtained
by calculating the eigenmodes of the matrix $K_{ij}/M$. The result
is $N$ eigenvalues $\omega_{m}^{2},\,\mbox{for }m=1,2,...,N$ (with
corresponding frequencies $\omega$).

\section{Limits to the validity of mean field theory }

\label{sec:limitsOfMF}

Here we calculate the domain over which a mean field theory (MF) treatment
of $\hat{J}_{z}^{2}$ is valid. In the special case of uniform Ising
coupling (e.g., $0<\mu_{R}-\omega_{1}\ll\omega_{1}-\omega_{2}$),
we have 
\begin{equation}
\hat{H}_{I}=\frac{2\chi}{N}\left(\sum_{i=1}^{N}\hat{\sigma}_{i}^{z}/2\right)\left(\sum_{j=1}^{N}\hat{\sigma}_{j}^{z}/2\right)=\frac{2\chi}{N}\hat{J}_{z}^{2},\label{eq:deriveJz2}
\end{equation}
where we have neglected a constant offset and $\hat{J}_{z}=\sum_{i=1}^{N}\hat{\sigma}_{i}^{z}/2$
is the $z$-component of the total composite spin of the system. The
interaction strength $\chi$ is independent of ion-ion separation
and is given by

\begin{equation}
\chi\approx\frac{F_{0}^{2}}{\hslash2M}\frac{1}{\mu_{R}^{2}-\omega_{1}^{2}}.\label{eq:JijNearCOM}
\end{equation}

We start each experiment with $N$ spins all prepared in the $\left|\uparrow\right\rangle $
state. Here we use the composite spin picture where this state is
labeled $\left|J=N/2,\: M_{J}=N/2\right\rangle $. (Formally $\left|J,M_{J}\right\rangle $
labels a state that is an eigenstate of $\hat{\vec{J}}^{2}$ and $\hat{J}_{z}$
with eigenvalues $J(J+1)$ and $M_{J}$, where $\hat{\vec{J}}=\sum_{i}\vec{\hat{\sigma}}_{i}/2$
is the total system spin.) The first pulse of the spin-precession
measurement sequence rotates the composite Bloch vector by an angle
$\theta_{1}$. After the rotation by $\theta_{1}$ the state can be
written as 
\begin{equation}
\left|\psi\right\rangle =\sum_{M_{J}=-N/2}^{N/2}C(J,\, M_{J})\left|J,\, M_{J}\right\rangle ,
\end{equation}
where the coefficients $C(J,\, M_{J})$ are significantly non-zero
for a small range $\Delta M_{J}$ centered on $M_{J}^{(0)}$, where
$M_{J}^{(0)}\approx\frac{N}{2}\cos\theta_{1}$. For a coherent spin
state, $\Delta M_{J}\lesssim\sqrt{N}$. 

\begin{flushleft}
We want to establish that a $\hat{J}_{z}^{2}$ interaction looks like
precession $\propto M_{J}^{(0)}$, at least over short durations.
A precession by an angle $\phi$ about the $z$-axis is obtained with
the operation $e^{-i\phi\hat{J}_{z}}$,
\begin{equation}
e^{-i\phi\hat{J}_{z}}\left|\psi\right\rangle =\sum_{M_{J}}C(J,\, M_{J})\, e^{-i\phi\hat{J}_{z}}\left|J,\, M_{J}\right\rangle \:.
\end{equation}
We measure expectation values of the total system spin $\hat{\overrightarrow{J}}$.
It is sufficient therefore to consider how these expectation values
transform under the rotation $e^{i\phi\hat{J}_{z}}$,
\begin{equation}
\begin{array}{ccc}
\left\langle \psi\right|\hat{J}_{y}\left|\psi\right\rangle  & \rightarrow & \left\langle \psi\right|e^{i\phi\hat{J}_{z}}\hat{J}_{y}e^{-i\phi\hat{J}_{z}}\left|\psi\right\rangle \end{array}.
\end{equation}
For simplicity, only matrix elements of the total spin in the $\hat{y}$-direction
$\hat{J}_{y}$ are considered. Identical expressions are also obtained
for $\hat{J}_{x}$.
\begin{equation}
\begin{array}{ccc}
\left\langle \psi\right|\hat{J}_{y}\left|\psi\right\rangle \\
=\sum_{M_{J}^{'},M_{J}}C(J,\, M_{J}^{'})^{\star}C(J,\, M_{J})\left\langle J,\, M_{J}^{'}\left|\hat{J}_{y}\right|J,\, M_{J}\right\rangle \\
=\sum_{M_{J}}\left\{ C(J,\, M_{J}-1)^{\star}C(J,\, M_{J})\left\langle J,\, M_{J}-1\right|\hat{J}_{y}\left|J,\, M_{J}\right\rangle \right.\\
\left.\quad\quad\quad+C(J,\, M_{J}+1)^{\star}C(J,\, M_{J})\left\langle J,\, M_{J}+1\right|\hat{J}_{y}\left|J,\, M_{J}\right\rangle \right\} .
\end{array}
\end{equation}
The double sum is eliminated by making use of the property that $\hat{J}_{y}$
has only non-zero matrix elements between states with $M_{J}^{'}-M_{J}=\pm1$.
$C(J,\: M_{J}\pm1)$ can be defined to be $0$ if $M_{J}-1=-N/2-1$
or $M_{J}+1=N/2+1$. Similarly
\begin{equation}
\begin{array}{ccc}
\left\langle \psi\right|e^{i\phi\hat{J}_{z}}\hat{J}_{y}e^{-i\phi\hat{J}_{z}}\left|\psi\right\rangle \\
=\sum_{M_{J}^{'},M_{J}}C(J,\, M_{J}^{'})^{\star}C(J,\, M_{J})e^{iM_{J}^{'}\phi}e^{-iM_{J}\phi}\left\langle J,\, M_{J}^{'}\left|\hat{J}_{y}\right|J,\, M_{J}\right\rangle \\
=\sum_{M_{J}}\left\{ C(J,\, M_{J}-1)^{\star}C(J,\, M_{J})e^{-i\phi}\left\langle J,\, M_{J}-1\right|\hat{J}_{y}\left|J,\, M_{J}\right\rangle \right.\\
\left.\quad\quad\quad+C(J,\, M_{J}+1)^{\star}C(J,\, M_{J})e^{i\phi}\left\langle J,\, M_{J}+1\right|\hat{J}_{y}\left|J,\, M_{J}\right\rangle \right\} \:.
\end{array}\label{eq:phi precession}
\end{equation}
This is to be compared with the same matrix element under the $H_{I}=\frac{2\chi}{N}\hat{J}_{z}^{2}$
interaction,
\begin{equation}
\begin{array}{ccc}
\left\langle \psi\right|e^{i\frac{2\chi}{N}\hat{J}_{z}^{2}t}\hat{J}_{y}e^{-i\frac{2\chi}{N}\hat{J}_{z}^{2}t}\left|\psi\right\rangle \\
=\sum_{M_{J}^{'},M_{J}}C(J,\, M_{J}^{'})^{\star}C(J,\, M_{J})e^{i\frac{2\chi}{N}\left(M_{J}^{'}\right)^{2}t}e^{-i\frac{2\chi}{N}\left(M_{J}\right)^{2}t}\left\langle J,\, M_{J}^{'}\left|\hat{J}_{y}\right|J,\, M_{J}\right\rangle \\
=\sum_{M_{J}}\left\{ C(J,\, M_{J}-1)^{\star}C(J,\, M_{J})e^{-i\frac{2\chi}{N}\left(2M_{J}-1\right)t}\left\langle J,\, M_{J}-1\right|\hat{J}_{y}\left|J,\, M_{J}\right\rangle \right.\\
\left.\quad\quad\quad+C(J,\, M_{J}+1)^{\star}C(J,\, M_{J})e^{i\frac{2\chi}{N}\left(2M_{J}+1\right)t}\left\langle J,\, M_{J}+1\right|\hat{J}_{y}\left|J,\, M_{J}\right\rangle \right\} \:.
\end{array}\label{eq:Jz2 matrix element}
\end{equation}
Equation \ref{eq:Jz2 matrix element} approximates a rotation about
the $\hat{z}$-axis (that is, approximates Eq. \ref{eq:phi precession})
if $\frac{2\chi}{N}t\left(2M_{j}-1\right)\approx\frac{2\chi}{N}t\left(2M_{j}+1\right)\approx\frac{2\chi}{N}t\cdot2M_{J}^{(0)}$
for all $M_{J}$ for which $C(J,\: M_{J})$ is significantly non-zero.
This is satisfied for short periods $t$ satisfying $\frac{2\chi}{N}t\cdot2\left(\Delta M_{J}\right)\ll1$.
For the initial coherent spin state, $\Delta M_{J}\lesssim\sqrt{N}$,
which puts a limit
\begin{equation}
\chi t\ll\frac{\sqrt{N}}{4}\label{eq:mean field time bound}
\end{equation}
on the period $t$ for which the $\hat{J}_{z}^{2}$ interaction can
be approximated as a precession $\propto M_{J}^{(0)}$. The precession
frequency predicted by the $\hat{J}_{z}^{2}$ analysis, $\frac{2\chi}{N}\cdot2M_{J}^{(0)}=\frac{2\chi}{N}\cdot2\frac{N}{2}\cos\theta=2\chi\cos\theta$,
is in agreement with the precession frequency predicted by the MF
analysis. 
\par\end{flushleft}

\begin{figure}
\begin{centering}
a)\includegraphics[width=0.6\columnwidth]{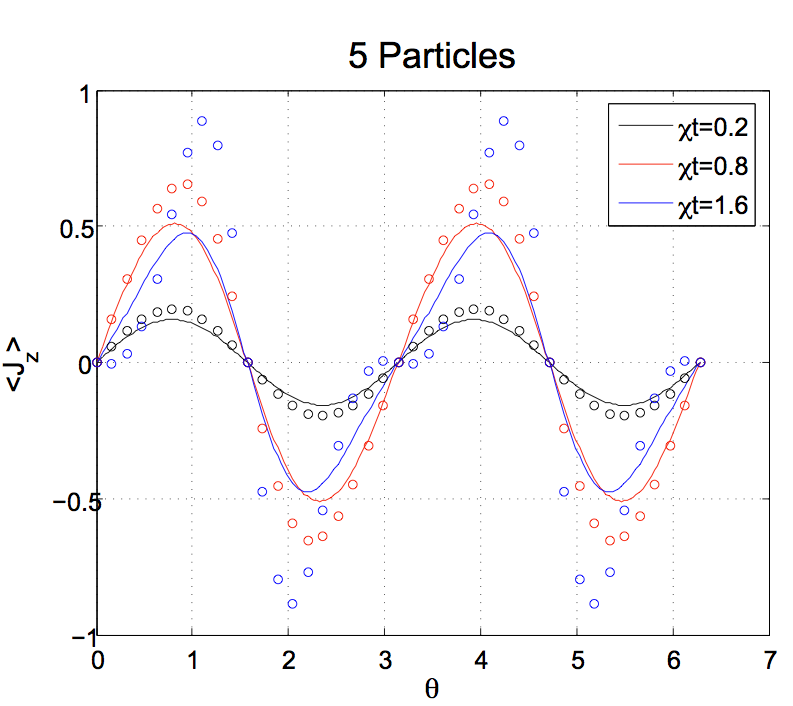}
\par\end{centering}

\begin{centering}
b)\includegraphics[width=0.6\columnwidth]{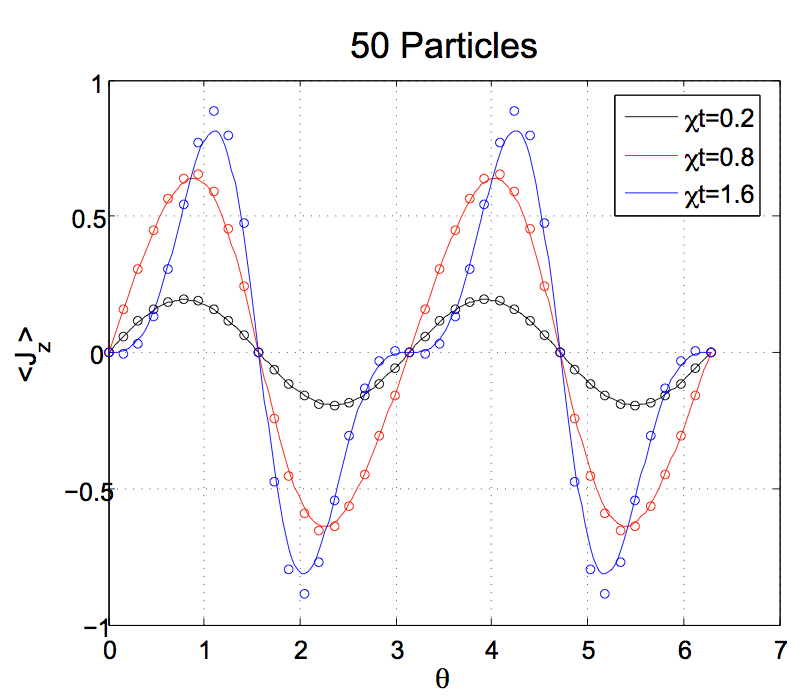}
\par\end{centering}

\begin{centering}
c)\includegraphics[width=0.6\columnwidth]{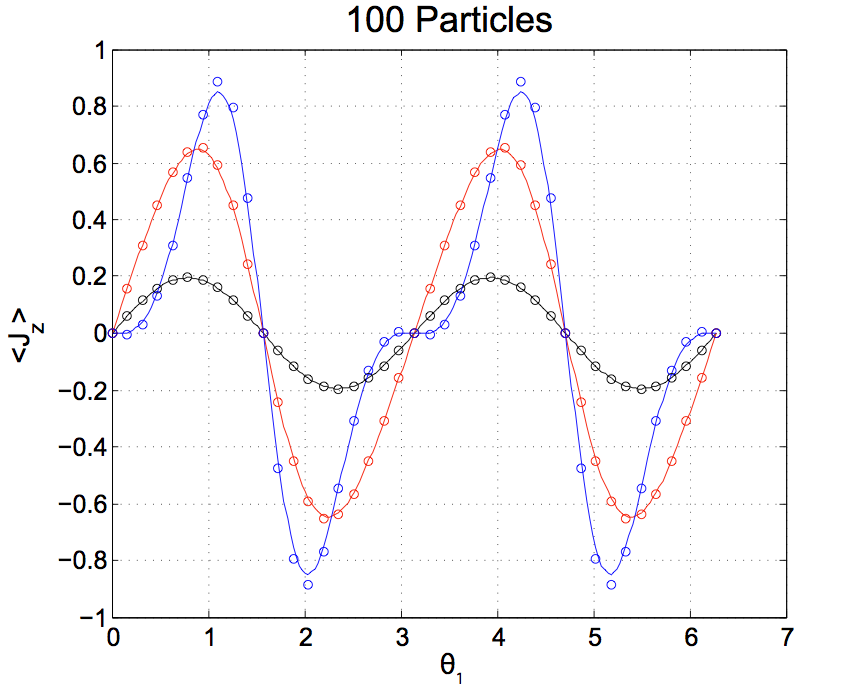}
\par\end{centering}

\caption{$\left\langle \hat{J}_{z}\right\rangle $ vs initial polar angle $\theta_{1}$
for an initial coherent spin state of $N$ spins. $t=2\tau_{arm}$
in the legend is the total interaction period. The solid line is the
exact calculation. The open circles are from the MF precession formula
(Eq.~\ref{eq:benchmarking}, with $\Gamma=0$). (a) $N=5$. (b) $N=50$.
(c) $N=100$.}

\label{fig:SM_Uys_Neq5} \label{fig:SM_Uys_Neq50} \label{fig:SM_Uys_Neq100}
\end{figure}

Equation \ref{eq:mean field time bound} provides a limit on the period
for which the MF analysis is valid. We confirm this limit through
explicit calculation of the spin precession benchmarking sequence
for the $\hat{H}_{I}=\frac{2\chi}{N}\hat{J}_{z}^{2}$ interaction
with $N=5$, $50$ and $100$ spins. The exact calculation is possible
because the system remains in the symmetric subspace (dimension of
Hilbert space is $2N+1$). Figure~\ref{fig:SM_Uys_Neq5}(a) shows
the results of the calculation for five spins. Reasonable agreement
between the MF precession formula (Eq.~\ref{eq:benchmarking} with
$\Gamma=0$) and the exact calculation is obtained for $\chi t=0.2\,$,
but not for $\chi t=0.8$ and $1.6$, as expected from Eq.~\ref{eq:mean field time bound}.
Figures~\ref{fig:SM_Uys_Neq50}(b) and \ref{fig:SM_Uys_Neq100}(c)
compare the exact calculation with the MF formula for $N=50$ and
$100$ spins. Excellent agreement between the MF formula and the exact
calculation is obtained for $\chi t=0.2$ and $0.8$. Some differences
are observed at $\chi t=1.6$. These differences decrease as $N$
increases.

\pagebreak{}

\bibliographystyle{apsrev4-1}
\bibliography{/Users/jwbritto/Documents/pinky/mendeley2/library}

\end{document}